\documentstyle[12pt]{article}
\catcode`\@=11

\catcode`\@12
\begin{document}
\newcommand{\eqn}[1]{(\ref{eq:#1})}
\newcommand{\ben}{\begin{equation}}
\newcommand{\een}{\end{equation}}
\newcommand{\bea}{\begin{eqnarray}}
\newcommand{\eea}{\end{eqnarray}}
\newcommand{\nn}{\nonumber \\ }
\newcommand{\bdm}{\begin{displaymath}}
\newcommand{\edm}{\end{displaymath}}
\newcommand{\A}{\cal{A}_\gamma }
\newcommand{\At}{\tilde{\cal{A}}_\gamma}
\newcommand{\Ak}{{\cal A}_{3k}}
\newcommand{\SU}{\widehat{SU(2)}}
\newcommand{\SUkp}{\widehat{SU(2)}_{\tilde{k}^+}}
\newcommand{\SP}{\widehat{SU(2)^+}}
\newcommand{\SM}{\widehat{SU(2)^-}}
\newcommand{\SSS}{\widehat{SU(3)}_{\tilde{k}^+}}
\newcommand{\YS}{Y^{SU(3)}}
\newcommand{\YW}{Y^{WS}}
\newcommand{\Zi}{{\sf Z}}
\newcommand{\HLk}{{\cal H}(\Lambda ,\ktp)}
\newcommand{\br}{\langle\!}
\newcommand{\kt}{\!\rangle}
\newcommand{\bra}[1]{\langle {#1}|}
\newcommand{\ket}[1]{|{#1}\rangle}
\newcommand{\lm}{\ell^-}
\newcommand{\lp}{\ell^+}
\newcommand{\la}{\lambda}
\newcommand{\LA}{\Lambda}
\newcommand{\al}{\alpha}
\newcommand{\eps}{\epsilon}
\newcommand{\va}{\vec{a}}
\newcommand{\vl}{\vec{\lambda}}
\newcommand{\oq}{\overline{q}}
\newcommand{\hf}{\frac{1}{2}}
\newcommand{\pa}{\partial}
\newcommand{\kp}{k^+}
\newcommand{\km}{k^-}
\newcommand{\ppu}{\psi^+}
\newcommand{\ppd}{\psi_+}
\newcommand{\pmu}{\psi^-}
\newcommand{\pmd}{\psi_-}
\newcommand{\ktp}{\tilde{k}^+}
\newcommand{\ktm}{\tilde{k}^-}
\newcommand{\com}{\marginpar{change}}
\begin{titlepage}
\begin{flushright}
DTP/92/49\\
NBI-HE-92-73\\
November  1992\\
\end{flushright}
\vskip 1.0cm
\begin{center}
{\Large\bf Coset Construction and Character Sumrules for \\ the
Doubly Extended N=4 Superconformal Algebras}\\
\vskip 1.cm
{\bf
{}~Jens~Lyng~Petersen$^a$\\
\vskip 1.cm
and\\
\vskip 1.cm
{}~Anne~Taormina$^{b}$}\\
\vskip 2cm
$^a$The Niels Bohr Institute, Blegdamsvej 17,\\
DK-2100 Copenhagen \O , Denmark.\\
\vskip 0.3cm
$^b$Department of Mathematics, University of Durham, South Road,\\
Durham, DH1 3LE, England.\\
\vskip 3mm
\end{center}
\vskip 1.0cm
\begin{center}
{\bf ABSTRACT}
\end{center}
\begin{quote}
Character sumrules associated with the realization of the $N=4$ superconformal
algebra $\At$ on manifolds corresponding to the group cosets
$SU(3)_{\ktp }/U(1)$ are derived and developed as an important tool
in obtaining the modular
properties of $\At$ characters as well as information on certain extensions
of that algebra. Their structure strongly suggests the existence
of rational conformal field theories with central charges in the range
$1 \le c\le 4$. The corresponding characters appear in the massive
sector of the sumrules and are completely specified in terms of the
characters for the parafermionic theory $SU(3)/(SU(2)\times U(1))$
and in terms of the branching
functions of massless $\At$ characters
into $SU(2)_{\ktp }\times SU(2)_1$ characters.
\end{quote}
\vfill
\end{titlepage}

\section{Introduction}
\setcounter{equation}{0}

The doubly extended $N=4$ superconformal algebras $\A$ are a one-parameter
family ($\gamma \in [\hf ,\infty[$) of linear superconformal algebras, i.e.
superconformal algebras containing a {\em finite dimensional} superalgebra.
They are characterized by their finite dimensional superalgebra being the
{\em non simple} ($D(2|1;\frac{\gamma }{1-\gamma })\oplus u(1)$),
and they contain $N=4$
supersymmetries, which is a maximum for linear superconformal algebras
{\cite{s89}}. $\A$ furthermore contains all linear conventional superconformal
algebras \cite{stvps88,stvp88} and provides a general framework to study
properties of these
subalgebras. Moreover, it has the challenging
feature of having no unitary representations falling in a minimal series
(for $k^->1$, see below).
However, non trivial physical applications where conformal invariance plays a
central role, such as superstring compactifications, are only well understood
when the theory is rational.
The Landau-Ginzburg technique developed for
describing  $N=2$, $c=9$ superstrings compactified to 4 dimensions is a famous
example of this.
It is therefore natural to seek  simple extensions of the chiral algebra
$\A$ which give rise to finite dimensional representations of the modular
group.
More precisely, we study extensions of
the (non-linear $N=4$) algebra $\At$, whose direct sum with the algebra
of four free fermions and one free boson $A^{Q,U}$  coincides with $\A$.
Hints for the existence of such extensions are found when analyzing
the character sumrules associated with the realization of $\At$ via coset
constructions based on quaternionic symmetric spaces,
which were discovered a few years ago \cite{vp,theo}.
It was suggested in \cite{opt} that extensions of $\At$ yield finite
dimensional representations of
the modular group. However, the full information on the character sumrules,
which is needed to completely identify the particular
rational extensions of $\At$ involved, was not available then.
In this paper, we will consider the coset $SU(3)_{\ktp} /U(1)$
which together with 4 free (Wolf space) fermions \cite{gptvp}
provides a realization of $\At$ for $\gamma = 2/(\ktp+3)$ and
derive the corresponding character sumrules
with particular emphasis on the massive sector.
In Section 2, we argue  that the functions $F^{\LA }_{2\lp ,n}(q)$ which
couple to the
$\At$ massive characters in the sumrules are related
both to the branching functions of $\At$ into its Kac-Moody subalgebra
$\SU_{\ktp} \times \SU_{1}$
and to the parafermions for the $SU(3)/(SU(2)\times U(1))$ theory.
They are argued to be labelled by a rational conformal theory at central charge
$c=c_{\phi }=1+3(\ktp -1)^2/(\ktp +1)(\ktp +3)$.
As a concrete example, the cases  $\gamma =1/2$
and $\gamma = 2/5$ are discussed in detail in  Appendix A.
The analytic structure
of the character sumrules associated to the coset constructions realizing
$\At$ is best understood when highest weight states of the $\At$
and of a certain rational
Gaussian model are explicitly constructed in terms of coset
operators. This is explained in Section 3, which indeed provides proofs
of some of the results alluded to in \cite{opt}.

The character sumrules also lead to the derivation of the modular properties
for the  massless characters of the $\At$ algebra and for the functions
$F^{\LA }_{2\lp ,n}(q)$, provided one assumes the decoupling of the
massless and massive sectors of the sumrules under modular transformations.
This hypothesis is justified in Section 4. The massless
characters and their modular transformations were presented in
\cite{opt}, but a detailed proof of these transformation properties
is given in Section 4 and in Appendix C.

\section{Analytic structure of the character sumrules }
\setcounter{equation}{0}

The $N=4$ superconformal algebra $\A$, whose affine subalgebra is
$\SU _{\kp} \times \SU_{\km} \times \widehat{U(1)}$, has been studied by
various groups over the last few years
\cite{stvps88,theo,gptvp,ivan,pt1,pt2,pt3,imp}.
It consists of the dimension 2 Virasoro generator $L(z)$,
seven dimension 1 currents
$T^{\pm i}(z)~~ (i=1,2,3)$ and $U(z)$, as well as four dimension $3/2$
supercurrents $G^a(z)$ and four dimension $1/2$ currents (free fermions)
$Q^a(z)$ which transform
as doublets under the two $\SU$ algebras. The commutation relations can be
found, for instance, in \cite{stvps88,stvp88,gptvp}. The algebra
is characterized by the levels
$\kp,\km$ (positive integers in the unitary representations
to which we shall restrict our attention)
of the two commuting $\SU$
subalgebras, or equivalently, by the parameter
$\gamma = \km/k$ and the central charge $c=6\kp \km/k$,
where $k=\kp +\km $. In fact,
the algebra $\A$ itself is non-simple, but is the
direct sum of the subalgebras, $\At$ and
${\cal A}^{QU}$: the algebra of free fermions and the affine $U(1)$.
The generators of $\At$, which is a non-linear superconformal algebra, are
constructed from those of $\A$ as explicitly shown in \cite{gs}
(see also \cite{gptvp}). Together
with a dimension 2
Virasoro generator $\tilde{L}(z)$, $\At$ contains six currents
$\tilde{T}^{\pm i}(z) (i=1,2,3)$ which are dimension
1 primaries wrt $\tilde{L}(z)$
and generate two $\SU$ algebras at level $\tilde{k}^{\pm }=k^{\pm }-1$, and
four dimension $3/2$ supercurrents $\tilde{G}^a(z)$. The non-linearity of the
algebra is encoded in the (anti)commutation relations of the dimension $3/2$
currents \cite{pt2}.

The representation theory and corresponding characters were given in
\cite{gptvp,pt1,pt2} for unitary highest weight state representations. These
are labelled  by the two isospin quantum
numbers $\lp ,\lm $, and have conformal
dimension $\tilde{h}$, whose lower bound $\tilde{h}_0$ is a function of
$\lp ,\lm ,\ktp $ and $\ktm $. An irreducible representation with
conformal dimension
$\tilde{h}_0$ is called {\em massless} or {\em chiral}
(it corresponds in the Ramond (R) sector to a
representation with non-zero Witten index), while any representation with
conformal weight $\tilde{h} > \tilde{h}_0$ is called {\em massive}. For fixed
$\ktp ,\ktm $, there is a finite number of massless and an
infinite number of massive representations and corresponding characters.
In the Neveu Schwarz (NS) sector for instance,
the massless characters of the $\At$ algebra are labelled  as
\ben
Ch_0^{\At, NS}(\ktp,\ktm,\lp ,\lm ,\tilde{h}_0^{NS};q,z_+,z_-)
\een
with $\ell ^{\pm }=0,\hf ,1,...,\frac{\tilde{k}^{\pm}}{2}$ and
\ben
\tilde{h}_0^{NS}=\frac{1}{k}[(\lp -\lm)^2+\kp \lm +\km \lp],
\label{eq:nsbound}
\een
while the massive characters are denoted by
\ben
Ch_m^{\At,NS}(\ktp,\ktm, \lp ,\lm ,\tilde{h}^{NS};q,z_+,z_-)
\een
with $\ell ^{\pm}=0,\hf ,...,\frac{\tilde{k}^{\pm }-1}{2}$ and
$\tilde{h}^{NS} > \tilde{h}_0^{NS}$. Often, for simplicity we shall suppress
several of the variables.

Realizations of $\At$ on manifolds based on group cosets of the form
$SU(\ktm+2)/(SU(\ktm)\times U(1))$
together with $4\ktm$ free (Wolf Space) fermions do
exist \cite{stvps88,vp,theo}.
The factor $U(1)$ in the denominator
precisely corresponds to the $U(1)$ current which decouples from $\A$
when considering the W-algebra $\At$, and the underlying quaternionic
symmetric space (or Wolf space) is $W=SU(\ktm +2)/(SU(\ktm )\times SU(2)
\times U(1))$.
The case $\ktm =0$ ($\ktp \in N^*$)
is very special and was analysed in \cite{pt3}: the
Wolf Space is empty, and $\At$ is realized by 3 $\SUkp$ currents and their
associated Sugawara form $\tilde{L}(z)$. The
$\At$ characters, which in this case
are bound to be massless, are affine $\SUkp$ characters
and the theory is rational.

The study of character sumrules corresponding to the above
coset realizations in the generic case $\ktm >0$ should shed some light on
the structure of some of the possible rational extensions of $\At$.
{}For much of this paper we shall
restrict our analysis to the value $\ktm =1$, for which the
group coset reduces to
$\SSS /U(1)$.
The cases $\ktm > 1$ are associated with
cosets $SU(\ktm +2)/(SU(\ktm)\times U(1))$ and certainly
merit further investigation\cite{vp,gptvp}. \footnote{Note that for $\ktm =1$,
the choice $\ktp =5$ leads to a contribution of 9
from the central charge of $\A$, which could be relevant in some new
compactification scheme.}

{}For a given value of $\ktp$, the realization is based
on a Hilbert space written as the direct product of the representation
space ${\cal{H}}_{\LA}^{SU(3)}$ for the affine Lie algebra $\SSS$
and corresponding highest weight $\LA=(\vec{a},\ktp,0)$ where
$\va =(a_1,a_2)$ is a highest weight in the Dynkin basis
of an $SU(3)$ representation (while
the last entry denotes the grade, an index we shall not use too much)
\cite{GeWi}, and the Fock space $\cal{H}^{WS}$
for the four free fermion fields associated with the Wolf space
$SU(3)/(SU(2)\times U(1))$. In addition to representations for $\At$,
the above
Hilbert space provides representations for the rational torus algebra
${\cal{A}}_{3k}$ (the extension of a $U(1)$ algebra by a dimension $3k$
operator, \cite{dijk})
as mentioned briefly in \cite{opt} and proven in detail in
sect. 3 ($k\equiv k^++k^-=\ktp +\ktm +2$).
There it is shown how the ``decoupling''
$U(1)$ generator of the algebra $\Ak$ emerges as the direct sum of the $U(1)$
hypercharge of $\SSS$ and the $U(1)$ generator of $SO(4)_1$.
The above information is encoded in the following
character sumrules, written here for the NS sector,
\ben
\chi ^{WS,NS} \cdot \chi_{\LA}=
\{ \chi^{WS,NS} \cdot \chi _{\LA}\}_0
+\{ \chi^{WS,NS} \cdot \chi _{\LA}\}_m,
\label{eq:charsum}
\een
where the character for the WS fermions is related to a reducible
$SO(4)_1$ representation,
\bea
\chi^{WS,NS}&\equiv& \chi^{WS,NS}(q,z_-,z_y)=
\frac{\theta_3(q,z_-z_y)}{\eta(q)}\cdot\frac{\theta_3(q,z_-/z_y)}
{\eta(q)}\nn
&=&\prod_{n=1}^{\infty}(1+z_-z_yq^{n-\hf })(1+z_-^{-1}z_y^{-1}q^{n-\hf })
(1+z_-z_y^{-1}q^{n-\hf })(1+z_-^{-1}z_yq^{n-\hf }),
\label{eq:WS}
\eea
and we denote the character for the unitary representation of
$SU(3)_{\ktp }$ with highest weight $\LA=(\vec{a},\ktp,0)$, by,
\ben
\chi_{\LA}\equiv \chi_{\LA}^{SU(3)_{\ktp }}(q,z_+,z_y).
\een
The variable $z_y$ is associated with the ``hypercharges", $\YS,\YW$ (see
next section). Unitarity requires the $SU(3)$ representations to be
integrable,
\ben
\ktp - <\!\!\vec{a},\psi \!\!>~ \ge 0
\label{eq:integrability}
\een
with $\psi $ the highest root in $SU(3)$, or $a_i$ non-negative
integers such that
$$0\le a_1+a_2 \le \ktp.$$

The RHS of \eqn{charsum} contains terms involving massless $\At$ characters
and terms involving massive $\At$ characters. However,
the massive representations reduce into two massless ones as
$\tilde{h}^{NS}$ reaches the lower bound \eqn{nsbound}
according to the formula \cite{pt2}
\ben
Ch_0^{\At,NS}(\lp ,0)+Ch_0^{\At,NS}\left(\lp+\hf,\hf\right)
=Ch_m^{\At,NS}(\lp ,\lm =0) \equiv Ch_m^{\At,NS}(\lp ).
\label{eq:relation}
\een
We recall that for $\ktm =1$, $2\lm =0,1$ in massless characters
and $2\lm =0$ in massive ones.
The splitting into massless and massive contributions in \eqn{charsum}
may therefore be considered non-unique.
{}For our purposes, the combinations of massless characters:
\bea
Ch_0^{\At,NS}(L=0)&\equiv&
-Ch_0^{\At, NS}\left(0,\hf\right),\nn
Ch_0^{\At,NS}(L=1,...,k-3)&\equiv&
\hf\left[Ch_0^{\At,NS}\left(\hf (L-1),0\right)
-Ch_0^{\At,NS}\left(\hf L,\hf \right)\right],\nn
Ch_0^{\At,NS}(L=k-2)&\equiv&
Ch_0^{\At,NS}\left(\hf(k-3),0\right),
\label{eq:combi}
\eea
introduced in \cite{opt} play a central role.
In fact the hypothesis for any
$\ktp $, that the massless and massive sectors defined according to these
combinations decouple from each other under modular transformations, is
extremely non-trivial, but is in fact
consistent with the modular properties of $\SSS$ and $\Ak$
characters. In particular, the massless combinations
will be shown to transform as $\widehat{SU(2)}_{\ktp +1}$
characters under modular transformations in Section 4. The equations
determining these modular transformations are tremendously overdetermined, and
the fact that the above decoupling assumption leads to a consistent solution
is at present our strongest evidence in favour of that assumption.
With this definition,
the massless part of the sumrule \eqn{charsum} is given by
\ben
\{\chi^{WS,NS}(q,z_y,z_-)\cdot\chi_{\LA}(q,z_y,z_+)\}_0\equiv
\sum_{L=0}^{k-2}{M_{\LA}}^L(q,z_y)Ch_0^{\At,NS}(L,q,z_+,z_-),
\label{eq:masslesssum}
\een
where the matrix ${M_{\LA}}^L$ (given in \cite{opt} without proof)
will be proven in the next section to be
\bea
\lefteqn{{M_{\LA}}^L(q,z_y)=}\nn
&&-\delta_{L,0}~\left\{\delta_{a_1,0}~\chi^{3k}_{-a_2+3(1+a_2)}(q,z_y)+
\delta_{a_2,0}~\chi^{3k}_{a_1-3(1+a_1)}(q,z_y)+
\delta_{a_1+a_2,k-3}~\chi^{3k}_{a_1-a_2+3k}(q,z_y)\right\}\nn
&&+\delta_{L,k-2}~\left\{\delta_{a_1,0}~\chi^{3k}_{-a_2-3(k-1-a_2)}(q,z_y)
+\delta_{a_2,0}~\chi^{3k}_{a_1+3(k-1-a_1)}(q,z_y)+
\delta_{a_1+a_2,k-3}~\chi^{3k}_{a_1-a_2}(q,z_y)\right\}\nn
&&+(1-\delta_{L,0})~(1-\delta_{L,k-2})~\biggl\{
\delta_{L,k-2-a_1}~\chi^{3k}_{a_1-a_2-3(k-1-a_2)}(q,z_y)\nn
&&+\delta_{L,k-2-a_2}~\chi^{3k}_{a_1-a_2+3(k-1-a_1)}(q,z_y)
+\delta_{L,a_1+a_2+1}~\chi^{3k}_{a_1-a_2}(q,z_y)
-\delta_{L,a_1}~\chi^{3k}_{a_1-a_2+3(1+a_2)}(q,z_y)\nn
&&-\delta_{L,a_2}~\chi^{3k}_{a_1-a_2-3(1+a_1)}(q,z_y)
-\delta_{L,k-3-(a_1+a_2)}~\chi^{3k}_{a_1-a_2+3k}(q,z_y)\biggr\},
\label{eq:matrix}
\eea
and
the characters $\chi_m^{3k}(q,z_y)$ appearing in the above matrix are those of
the rational torus algebra ${\cal{A}}_{3k}$. They are related to generalised
theta functions in the following way,
\ben
\chi_m^{3k}(q,z_y)=
\frac{1}{\eta(q)}\theta_{m,3k}(q,z_y^{2/3})
\equiv \frac{1}{\eta(q)}\sum_{n \in Z+\frac{m}{6k}}q^{3kn^2}z_y^{2kn}.
\label{eq:rational}
\een

Concerning the massive part of the sumrule we note  that
the massive characters (in the NS sector here) may be written quite generally
as
\cite{pt1,pt2},
\bea
Ch^{{\At},NS}_m(\ktp ,\ktm ,\lp,\lm,\tilde{h};q,z_+,z_-)&=&
q^{\tilde{h}-\tilde{h}_m}Ch^{{\At},NS}_m(\ktp ,\ktm ,\lp,\lm,\tilde{h}_m;
q,z_+,z_-)
\nn
&=&\chi_{c_\phi ,h_\phi}(q) \chi_F^{NS}(q,z_+,z_-)\chi^{\ktp -1}_{2\lp}(q,z_+)
\chi^{\ktm -1}_{2\lm}(q,z_-).\nn
\label{eq:massivefact}
\eea
Here, $\chi^{NS}_F(q,z_+,z_-)$ is a character for 4
free fermions (two complex NS ones), $\chi^{k}_{2\ell}$ is
the $\widehat{SU(2)}_k$ characters for isospin $\ell $ given by
\bea
\chi _{2\ell }^k(q,z)&=& q^{-\frac{1}{8}}z^{-1}
\prod _{n=1}^{\infty} (1-q^n)^{-1}(1-q^nz^2)^{-1}
(1-q^{n-1}z^{-2})^{-1}\nn
&\times & \sum _{m \in Z+\frac{1}{k+2}(\ell +\hf )}
q^{(k+2)m^2}[z^{2(k+2)m}-z^{-2(k+2)m}],
\eea
and finally
\ben
\chi_{c_\phi ,h_\phi}(q)=\frac{q^{h_\phi-(c_\phi -1)/24}}{\eta(q)}
\een
is the Virasoro character corresponding to
\bea
h_\phi&=&\tilde{h}-\tilde{h}_0+\delta \tilde{h}\nn
\delta\tilde{h}&\equiv&\frac{k^+k^-}{4k}\left (\frac{2\lp}{k^+}+
\frac{2\lm}{k^-}\right )\left (\frac{\ktp -1-2\lp}{k^+}+
\frac{\ktm -1-2\lm}{k^-}\right )\ge 0\nn
c_\phi&=&1+6\left (\sqrt{\frac{k^+k^-}{k}}-\sqrt{\frac{k}{k^+k^-}}\right )^2
\label{eq:central}
\eea
at conformal dimension $h_\phi$ and central charge $c_\phi$
with
$$\tilde{h}-\tilde{h}_m\equiv h_\phi-\frac{1}{24}(c_\phi -1).$$
Thus
\ben
\tilde{h}_m\equiv \tilde{h}_0+\frac{k^+}{2k}[\frac{2\lp +1}{k^+}-\hf]^2
\label{eq:hm}
\een
(for $k^-=\ktm+1=2$).
The interpretation of this structure in terms of free fields was recently
given in ref.~\cite{imp}. Then also
\ben
Ch^{{\At},NS}_m(\ktp ,\ktm ,\lp,\lm,\tilde{h};q,z_+,z_-)=
\chi_{c_\phi ,h_\phi}(q)\cdot\eta(q)\cdot
Ch^{{\At},NS}_m(\ktp ,\ktm ,\lp,\lm,\tilde{h}_m;q,z_+,z_-).
\een
{}For $\ktm =1$ only $\lm =0$ occurs in the massive case. We may then write the
massive part of the sumrule as
\ben
\{ \chi^{WS,NS}(q,z_y,z_-)\cdot\chi_{\LA}(q,z_y,z_+)\}_m=
\sum_{2\lp =0}^{\ktp-1} \sum_{n\in{\sf Z}_k}
\tilde{M}^{\LA}_{2\lp ,n}(q,z_+ ,z_-,z_y)
F^{\LA}_{2\lp ,n}(q)
\label{eq:massivesum}
\een
where
\ben
\tilde{M}^{\LA}_{2\lp ,n}(q,z_+ ,z_-,z_y)
=Ch^{{\At},NS}_m(\tilde{h}_m,\lp,0;q,z_+,z_-)\chi^{3k}_{-2(a_1-a_2)+6\lp +6n}
(q,z_y).
\label{eq:matrixt}
\een
This result will be proven in the next section.
We see that the functions $F_{2\lp ,n}^{\LA}(q)$ may be interpreted as an
infinite
sum of Virasoro characters, multiplied by $\eta$.
Together with the $\At$ character at the particular
value of conformal dimension given by $\tilde{h}_m$, they provide the
contribution
from an infinite sum of massive characters.
It follows from the structure of the sum rule  that these functions have the
form of a certain fractional power of $q$ multiplied by
an infinite power series  in $q$ with positive integer coefficients, which we
have  analyzed by algebraic manipulation programs for  $\ktp =1,2$.
{}From the strongly motivated assumption
(see sect. 4) that the massless and massive parts of the
sumrule decouple from each other under modular
transformations, it follows that
the functions $F^{\LA}_{2\lp ,n}(q)$ in fact also carry a finite dimensional
representation of the modular group. Furthermore, from the above discussion
it then further appears plausible
that they are related to characters of some extended algebra presumably
in fact giving rise to a rational conformal field theory at
$c=c_{\phi}$, perhaps corresponding to some $W$-extension of the Virasoro
algebra or some coset construction.

The modular forms
$F^{\LA}_{2\lp ,n}(q)$ satisfy the following symmetry relations:
\bea
F^{\LA }_{2\lp ,n}(q)&=&F^{\LA }_{2\lp ,n+k}(q),\nn
F^{\LA }_{2\lp ,n}(q)&=&F^{\LA _C}_{2\lp ,-n-2\lp}(q),
\label{eq:Fsym1}
\eea
as is obtained by letting $z_y \rightarrow z_y^{-1}$ in \eqn{charsum},
a shift which relates the sumrule for an integrable $\SSS$ representation
$\va =(a_1,a_2)$ to the sumrule for its conjugate $\va _C=(a_2,a_1)$. This
can be seen from the following properties of the $z_y$-dependent functions
in the sumrule,
\bea
\chi^{\SSS}_{\LA }(q,z_+,z_y^{-1})&=&\chi^{\SSS}_{\LA _C}(q,z_+,z_y),\nn
\chi^{WS,NS}(q,z_-,z_y^{-1})&=&\chi^{WS,NS}(q,z_-,z_y)
\label{eq:su3sym1}
\eea
and
\ben
\chi_m^{3k}(q,z_y^{-1})=\chi^{3k}_{-m}(q,z_y),
\een
with $\Lambda_C\equiv (\va_C,\ktp ,0)$.
Two other symmetries follow from allowing the sumrule \eqn{charsum} to
``flow'' as
\ben
z_\pm \rightarrow q^{\mp\hf} z_\pm ,~~z_y \rightarrow q^{3/2} z_y,
\een
or
\ben
z_\pm \rightarrow z_\pm ,~~z_y \rightarrow q^{\pm 1} z_y.
\label{eq:flow2}\een
Under the first flow, the $\SSS$ and the 4 fermion characters respectively
transform as
\bea
\chi^{\SSS}_{\LA }(q,q^{-\hf }z_+,q^{\frac{3}{2}}z_y)&=&
q^{-\ktp }(z_y/z_+)^{-\ktp }\chi^{\SSS}_{\LA }(q,z_+,z_y)\nn
\chi^{WS,NS}(q,q^{\hf }z_-,q^{\frac{3}{2}}z_y)&=&
q^{-\frac{5}{2}}z_-^{-1}z_y^{-3}\chi^{WS,NS}(q,z_-,z_y).
\label{eq:su3sym2}
\eea
On the other hand, considerations of spectral flow for $\At$ characters
\cite{defev,pt2} lead to
\bea
Ch^{\At, NS}_0(L,q,q^{-\hf }z_+,q^{\hf }z_-)&=&
-q^{-\frac{\ktp +1}{4}}z_-^{-1}z_+^{\ktp }
Ch^{\At, NS}_0(\ktp -L+1,q,z_+,z_-),\nn
Ch^{\At, NS}_m(\lp ,\lm =0,q, q^{-\hf }z_+,q^{\hf }z_-)&=&
q^{-\frac{\ktp +1}{4}}z_-^{-1}z_+^{\ktp } Ch^{\At, NS}_m(\frac{\ktp -1}{2}-\lp,
0,q,z_+,z_-),\nn
\eea
for $L=0,...,\ktp+\ktm$ (with $\ktm =1$ here). Here both sides are evaluated
for the value of confomal dimension equal to $h_m$
(eq.(\ref{eq:hm})).
Finally,
\ben
\chi_m^{3k}(q,q^{3/2}z_y)=q^{-3k/4}z_y^{-k}\chi_{m+3k}^{3k}(q,z_y).
\een
It is now straightforward to conclude that the modular forms
$F^{\LA }_{2\lp ,n}(q)$
obey the following symmetry relation,
\ben
F^{\LA }_{2\lp ,n}(q)=F^{\LA }_{\ktp-1 -2\lp ,n+2\lp +2}(q).
\label{eq:Fsym2}\een
The second flow \eqn{flow2} relates the character sumrules for
$\SSS$ representations
pertaining to the same orbit under the order 3 transformation
\ben
\phi (a_1,a_2)=(a_2,\ktp -(a_1+a_2)),
\label{eq:phi}
\een
\ben
\chi^{\SSS}_{\LA }(q,z_+,q^{\epsilon }z_y)=q^{-\frac{\ktp }{3}}
z_y^{-\frac{2\ktp }{3} \epsilon }
\chi^{\SSS}_{\phi ^{\epsilon }(\LA )}(q,z_+,z_y).
\label{eq:su3sym3}
\een
($\phi^{\epsilon}(\Lambda )\equiv (\phi^{\epsilon}(\va),\ktp ,0)$).
Furthermore, the 4-fermion character and the theta functions transform as
($\epsilon =\pm 1$)
\bea
\chi^{WS,NS}(q,z_-,q^{\epsilon }z_y)&=&q^{-1}z_y^{-2\epsilon }
\chi^{WS,NS}(q,z_-,z_y)\nn
\chi^{3k}_m(q,q^{\epsilon }z_y)&=&
q^{-\frac{k}{3}}z_y^{-\frac{2k}{3}\epsilon }\chi^{3k}_{m+2\epsilon k}(q,z_y).
\eea
These properties lead to the third relation between the modular forms
$F^{\LA }_{2\lp ,n}(q)$,
\ben
F^{\LA }_{2\lp ,n}(q)=
F^{\phi^{\epsilon }(\LA )}_{2\lp ,n+\epsilon +\hf (\epsilon -1)a_1+\hf
(\epsilon +1)a_2}(q).
\label{eq:Fsym3}
\een
The three sets of symmetries described so far allow for a complete
determination of the functions $F^{\LA }_{2\lp ,n}(q)$ in the very
particular case where $\ktp =1$. It is shown in Appendix A
that the massive sector of the sumrules in this case depends on the two
functions
\bea
2F^{((0,0),1,0)}_{0,0}(q)&=&\theta_{0,1}(q)\nn
2F^{((0,0),1,0)}_{0,1}(q)&=&\theta_{1,1}(q),
\eea
which, when multiplied by $1/\eta $, provide the two characters for the
$SU(2)_1$
conformal field theory
with central charge 1, in total agreement with the expected
value of $c_{\phi }$ \eqn{central}. When $\ktp \ge 2$, some further
information on the functions $F^{\LA }_{2\lp,n}(q)$ can be obtained by
deriving their modular properties as motivated in section 4, and by relating
them to the branching functions in the
reduction of $\At$ representations according to
their $\SU_{\ktp}\times\SU_{\ktm}$ content.
To this end it is useful to
separate the variables $z_+,z_-$ and $z_y$ in the sumrule \eqn{charsum}.
The character corresponding to four free fermions
may always be written in terms of $\widehat{SU(2)}_1$
characters in the following way (see also \cite{et2}),
\ben
\chi^{WS,NS}(q,z_-,z_y)=\chi^1_0(q,z_-)\chi^1_0(q,z_y)+\chi^1_1(q,z_-)
\chi^1_1(q,z_y).
\een

Next, the decomposition of the $\SSS$ characters in $\SUkp$ characters for
a regular embedding of
$SU(2)$ in $SU(3)$, is given by the
following general structure,
\ben
\chi^{\SSS}_{\LA }(q,z_+,z_y)=
\sum_{2\lp =0}^{\ktp }
\sum_{n \in Z_{\ktp }} P_{2\lp ,n}^{\LA}(q)
\chi_{4(a_1-a_2)+6(n+\lp )}^{3\ktp }(q,z_y)  \chi^{\ktp }_{2\lp }(q,z_+),
\een
where the functions $P_{2\lp ,n }^{\LA}(q)$
satisfy the periodicity condition
$P_{2\lp ,n+\ktp }^{\LA}(q)=P_{2\lp ,n}^{\LA}(q)$, and obey
the symmetry relations
\bea
P_{2\lp ,n}^{\LA _C}(q)&=&P_{2\lp ,-n-2\lp }^{\LA}(q),\nn
P_{2\lp ,n}^{\LA}(q)&=&P_{\ktp -2\lp ,n+2\lp }^{\LA}(q),\nn
P_{2\lp ,n}^{\LA}(q)&=&
P_{2\lp ,n-(\epsilon -1)
a_1-(\epsilon +1)a_2 }^{\phi^{\epsilon }(\LA)}(q),
\label{eq:Psym}
\eea
which immediately follow from the relations \eqn{su3sym1}, \eqn{su3sym2} and
\eqn{su3sym3}.

As discussed in \cite{Huitu}, the functions $P_{2\lp ,n}^{\LA}(q)$
actually provide the characters of the \break parafermionic theory
$SU(3)/(SU(2)\times U(1))$.
We illustrate in Appendix A  how to derive expressions for these
characters for the cases
$\ktp =1$ and $\ktp =2$ in
terms of familiar modular forms. For instance, in the simplest case where
$\ktp =1$, one has, for all $\va $, $\vec{a}'$ corresponding to integrable
representations
\eqn{integrability},
\ben
P_{0,0}^{(\va ,1,0)}(q)=P_{1,0}^{(\vec{a}',1,0)}(q)=1.
\een
The LHS of the sumrule \eqn{charsum} can now be rewritten as
\bea
\lefteqn{
\chi^{WS,NS}(q,z_-,z_y)\chi^{\SSS}_{\LA }(q,z_+,z_y)}\nn
&=&\sum_{2\lm =0}^1  \sum_{2\lp =0}^{\ktp }
\chi^{\ktp }_{2\lp }(q,z_+) \chi^1_{2\lm }(q,z_-)~
\sum_{n=0}^{\ktp -1} P_{2\lp ,n }^{\LA}(q)
\chi_{4(a_1-a_2)+6(n+\lp )}^{3\ktp }(q,z_y)
\chi^1_{2\lm }(q,z_y)\nn
&=&
\sum_{2\lm =0}^1  \sum_{2\lp =0}^{\ktp }
\chi^{\ktp }_{2\lp }(q,z_+) \chi^1_{2\lm }(q,z_-)
\sum_{n=0}^{\ktp -1} P_{2\lp ,n }^{\LA}(q)\\
&\times&
\sum_{r=0}^{k-1}
\frac{1}{\eta(q)}\theta_{2(\lm +r)\ktp -4(a_1-a_2)-6(n+\lp ),k\ktp }(q)
\chi_{6(\lm +r)+4(a_1-a_2)+6(n+\lp )}^{3k} (q,z_y).\nonumber
\label{eq:lhssum}
\eea
The last equality is based on the fact that the $\widehat{SU(2)}_1$
characters are related to theta functions at level 1
\ben
\chi^1_{2\lm }(q,z_y)\equiv \frac{1}{\eta(q)}\theta_{2\lm ,1}(q,z_y^2),
\een
and on another remarkable identity between theta functions, proven in
Appendix B,
\ben
\theta_{a,1}(q,z_y^2)\theta_{b,3\ktp }(q,z_y^{2/3})=
\sum_{r=0}^{k-1} \theta_{(a+2r)\ktp -b,k\ktp }(q)
\theta_{6r+3a+b,3k}(q,z_y^{2/3}).
\label{eq:identity}
\een
It is now clear from \eqn{lhssum}
that the $z_y$ dependence of the character sumrule is
completely encoded in the characters for the $\Ak$ algebra which were
given above \eqn{rational}. Indeed it follows that the realization of $\At$
involves that algebra as will be shown more explicitly in the next section.
Furthermore, the dependence on the variables
$z_+,z_-$ appears through the product of two $\SU$ characters, at levels
$\ktp $ and 1. In order to compare with the RHS of the sumrule \eqn{charsum},
one is naturally led to express the $\At$ characters in terms of the
characters for the $\SUkp \times \widehat{SU(2)_1}$ Kac-Moody subalgebra.
This is a relatively easy task for the massive $\At$ characters. Indeed, as
explained above, the
latter factorize in the product containing two affine $\SU$ characters at
level
$\ktp-1$ and $\ktm -1$ \cite{pt1}.
{}From this expression and from the GKO formula \cite{GKO}
\ben
\chi^{\ktp -1}_{2\lp}(q,z)\chi^1_{2\lm}(q,z)=
\sum^{\ktp}_{\stackrel{2\ell =0}{2\ell
\equiv 2\lp +2\lm \ \mbox{mod}\  2}}
\chi^{\ktp}_{2\ell}(q,z)\chi^{Vir,(\ktp +1)}_{2\lp +1,2\ell +1}(q)
\label{eq:gko}
\een
where $\chi^{Vir,(m)}_{p,q}$ are the unitary minimal model
Virasoro characters for $c=1-6/m(m+1)$, one easily works out that
\bea
\lefteqn{\eta(q)Ch_m^{\At,NS}(\ktp,\ktm,\tilde{h}^{NS}_m,\lp,\lm ;q,z_+,z_-)
}\nn
&=&
\sum^{\ktp}_{\stackrel{2\overline{\lp}=0}{}}
\sum^{\ktm}_{\stackrel{2\overline{\lm} =0}{\hspace{-1.3cm}
2\overline{\lp}+2\overline{\lm}
\equiv 2\lp +2\lm \ \mbox{mod}\  2}}
\chi_{2\overline{\lp} }^{\ktp}(q,z_+)\chi_{2\overline{\lm}}^{\ktm}(q,z_-)
\chi^{Vir,(\kp)}_{2\lp +1,2\overline{\lp}+1}(q)
\chi^{Vir,(\km)}_{2\lm +1,2\overline{\lm}+1}(q),
\label{eq:nsmassive}
\eea
where $\chi^{Vir,(k^{\pm })}_{2\ell ,2\overline{\ell} }(q)$
are Virasoro characters at
level $k^{\pm }$, and are understood to be 1 whenever $\kp$ or $\km$ are
integers smaller than 3.\

Similar expressions hold in the Ramond sector
and can be obtained by spectral flow from \eqn{nsmassive}.

{}For the massless characters it is not possible to provide similar explicit
expressions for the corresponding branching functions. However, these
branching
functions are related in a very interesting way to the modular forms
associated
with the massive sumrule above.
Let us introduce
the branching functions $Y^{(\ktp +2)}_{r,s}(q)$ for the NS $\At$ massless
characters when $\ktm =1$, so that the combinations \eqn{combi} are
written as
\ben
Ch_0^{\At,NS}(L,q,z_+,z_-)=
\sum^{\ktp}_{\stackrel{2\lp=0}{}}
\sum^{1}_{\stackrel{2\lm =0}
{\hspace{-1.3cm}2\lp+2\lm
\equiv L-1\ \mbox{mod}\  2}}
(-1)^{2\lp +L+1}Y^{(\ktp +2)}_{2\lp +1,L+1}(q)
\chi^{\ktp }_{2\lp }(q,z_+)\chi^1_{2\lm }(q,z_-).
\label{eq:branching}
\een
These branching functions seem to have an intriguing relation to
minimal unitary Virasoro characters \cite{next}.
There are as many functions $Y^{(\ktp +2)}_{m,n}(q)$ at fixed value of
$\ktp $ as there are Virasoro characters at level $\ktp +2$, they
possess the same symmetries as the corresponding Virasoro characters,
and they transform in a similar way under modular transformations \cite{next}.

Let us finish this section by summarizing the analytic structure of the
character sumrules in the NS sector when $\ktm =1$,
\bea
&&\sum_{2\lm =0}^1 \sum_{2\lp =0}^{\ktp }
\chi^{\ktp }_{2\lp }(q,z_+) \chi^1_{2\lm }(q,z_-)\nn
&\times&\sum_{n=0}^{\ktp -1} P_{2\lp ,n }^{\LA}(q)\sum_{r=0}^{k-1}
\frac{1}{\eta(q)}
\theta_{2(\lm +r)\ktp-4(a_1-a_2)-6(n+\lp ),k\ktp }(q)
\chi^{3k}_{6(\lm +r)+4(a_1-a_2)+6(n+\lp )} (q,z_y)\nn
&=&\sum_{2\lm =0}^1 \sum_{2\lp =0}^{\ktp }
\chi^{\ktp}_{2\lp }(q,z_+)\chi^1_{2\lm }(q,z_-)\nn
&\times &\sum_{\stackrel{L=0}{L+1\equiv 2\lp+2\lm\ \mbox{mod}\  2}}^{\ktp+1}
\{ (-1)^{2\lp +L+1}M_{\LA}^{~L}(q,z_y)Y_{2\lp +1,L+1}^{(\ktp +2)}(q)\nn
&+&(1-\delta _{L,0})(1-\delta_{L,k-2})
\sum_{n\in Z_k} \chi^{3k}_{-2(a_1-a_2)+6n+3(L-1)}(q,z_y)
F_{L-1,n}^{\LA}(q)\eta(q)^{-1}
\chi^{Vir,(\kp )}_{L,2\lp +1}(q)\}.\nn
&&
\label{eq:summary}
\eea

The information encoded in the above expression for the sumrules
and the symmetry properties \eqn{Fsym1}, \eqn{Fsym2}, \eqn{Fsym3},
\eqn{Psym} allow for
a complete determination of the functions $P_{2\lp ,n }^{\LA}(q)$,
and for expressing the functions $F_{2\lp ,n}^{\LA}(q)$
in terms of the
branching functions $Y^{(\ktp +2)}_{r,s}(q)$, as derived in Appendix A
for
$\ktp =1,2$. The latter are   not yet known for
$\ktp >1$,
and will be discussed further elsewhere \cite{next}. Let us however
mention that the results obtained by Kazama and Susuki \cite{KS} may
provide an
indirect route to the determination of these branching functions.
Indeed, these authors constructed
$N=2$ superconformal theories by considering cosets
of the form
\ben
\frac{SU(n+m)\times SO(2nm)}{SU(n)\times SU(m)\times U(1)}
\een
involving hermitian symmetric spaces. When $n=1$ and $m=2$, one has the
following schematic structure
\ben
SU(3)_{\ktp }\times SO(4)_1 \sim [N=2]_{\mbox{KS}}
\times SU(2)_{\ktp +1}\times U(1).
\een

On the other hand, our analysis gives

\ben
SU(3)_{\ktp }\times SO(4)_1 \sim [\At ]\times U(1),
\een
where we have found that both the $U(1)$ and the $[\At]$ occur with extended
algebras. This result
underlines once more \cite{theo} the intimate connection
between the doubly
extended $N=4$ superconformal algebras and these Kazama-Suzuki $N=2$ theories.

\section{Embedding of $\At\oplus\Ak$ in an affine $\SSS$ Module $\otimes$
Fermion Fock Space.}
\setcounter{equation}{0}

The reduction of this module according to the rational torus algebra
$\Ak$ and $\At$ was qualitatively described above.
It plays a crucial role for the derivation of the
results we present. In this section we indicate some details of the pertinent
proofs.

\subsection{Proof that $\At$ highest weight states come as multiplets forming
representations of $\Ak$}

{}For more details concerning the construction of $\At$
in terms of Wolf space fermions
and affine $\SSS$ we refer to ref.{\cite{gptvp}}. Here we briefly introduce
the
notation. The construction is based on two complex fermions,
$\ppu (z),\ppd (z)$
and $\pmu (z),\pmd(z)$ pairwise conjugate, as well as on the
affine currents of
$\SSS$, denoted $V^i(z),V_i(z)$  with $i\in\{+,-,\theta,m\}$.
The $\SUkp$ subalgebra of $\At$ is then generated by
$V^\theta=T^{++},V_\theta=T^{+-}$ and their
OPE, involving $T^{+3}$ given as
a combination of the Cartan generator currents, $V^m,V_m$.
Let us denote by $\YS$ the $U(1)$ current in $\SSS$ corresponding to the
usual $SU(3)$ ``hypercharge". It is proportional to
the combination of $V^m,V_m$ orthogonal to $T^{+3}$. Then $(V^+,V_-)$ form an
$\lp =\hf$ doublet with $\YS=+1$ and $(V^-,V_+)$ form an $\lp =\hf$ doublet
with $\YS=-1$.

The WS fermions build the $\widehat{SU(2)}^-_1$ generators as follows
\ben
T^{-+}=-2i\ppd\pmd ,\ \  T^{--}=-2i\ppu\pmu ,\ \ T^{-3}=\ppu\ppd+\pmu\pmd.
\een
So $(\pmd ,\ppu)$ form an $\lm =\hf$ doublet with $\YW =+1$ whereas
$(\ppd ,\pmu )$ form an $\lm =\hf$ doublet with $\YW =-1$. Here the $U(1)$
current corresponding to ``Wolf space hypercharge" is expressed by
$$Y^{WS}\equiv (\ppu\ppd -\pmu\pmd ).$$
The $U(1)$ current, $U(z)$ of $\A$ which decouples from $\At$ is then given by
\ben
U=i\frac{\sqrt{3}}{2} Y\equiv i\frac{\sqrt{3}}{2}(\YS+\YW ).
\een
The four supersymmetry currents of $\At$ are given by
\bea
G_+=\frac{1}{\sqrt{2}}(V^+\ppd +V^-\pmd )&&G_-=\frac{1}{\sqrt{2}}(V_+\ppu +
V_-\pmu ),\nn
G_{+K}=\frac{i}{\sqrt{2}}(V^+\pmu -V^-\ppu )&& G_{-K}=-\frac{i}{\sqrt{2}}
(V_+\pmd -V_-\ppd).
\eea
Finally the Virasoro generator of $\At$ may be expressed as (normal
ordering implied)
\ben
L=-\frac{1}{8}(V^iV_i+V_iV^i)-(\pa\ppd\ppu +\pa\ppu\ppd +\pa \pmd\pmu +
\pa\pmu\pmd)+\frac{1}{k}UU.
\een
The rational torus model character, \eqn{rational}, corresponds to
an irreducible representation of $\Ak$ labelled by the integer, $m$,
which splits into infinitely many
representations of $Y(z)=\frac{2}{i\sqrt{3}}U(z)$ with zero modes
$Y\equiv\frac{2}{\sqrt{3}}u$ each of which has
\ben
h^Y=\frac{3}{4k}Y^2=\frac{u^2}{k},~~~~~Y=\frac{1}{3}(m +6kn), \ n\in{\sf Z}.
\een
In ref.~{\cite{gptvp}} a finite set of $\At$ highest weight states were
explicitly
constructed in terms of the WS fermions and the $\SSS$ generators.
Massless as well as massive examples were presented.
Our present assertion is that these were merely examples of what are in fact
infinite families of $\At$ highest weight states in the space
$${\cal H}(\Lambda ,\ktp)={\cal H}^{WS}\otimes{\cal H}^{\SSS}_\Lambda ,$$
the tensor product of the Fock space for the WS fermions and a module for
$\SSS$ in the representation labelled, $\Lambda=(\va ,\ktp ,0)$, $\va$
being the weight of the corresponding $SU(3)$ representation.

In fact we shall now present an operator, $K(p)$ for all $p\in{\sf Z}$
with the
property that if $\ket{hws}$ is a certain  $\At$ highest weight state in
${\cal H}(\Lambda ,\ktp)$, then the set of states
$$\{\ket{hws,p}\equiv K(p)\ket{hws}|p\in{\sf Z}\}$$
have the following properties,
\begin{enumerate}
\item
they are $\At$ highest weight states;
\item
they are $Y$ highest weight states;
\item
they carry an irreducible representation of $\Ak$ labelled by an integer, $m$
easily computable in terms of $\ket{hws}$ (see below).
\end{enumerate}
{}For simplicity let us first consider the simplest possible $\At$ highest
weight
state; the WS vacuum tensored with the $\SSS$ singlet highest weight state.
Denote that combined state by $\ket{0}$. We define the action of $K(p)$ by
the state $\ket{p}$ obtained by
\ben
\ket{p}\equiv K(p)\ket{0}\equiv \left\{
\begin{array}{ll}
\{(V^+V_-)^{\ktp}(\ppu\pmd)^3\}^p & p\ge 0 \\
\{(V_+V^-)^{\ktp}(\ppd\pmu)^3\}^{|p|} & p< 0.
\end{array}
\right.
\een
Here we have introduced the following condensed notation: Let $\{A_i(z)\}$
be any set of primary field operators with some standard mode expansion. By
the state
$$A_{i_1}A_{i_2}\ldots A_{i_n}\ket{0},$$
we mean to first consider the state
$$A_{i_1}(z_1)A_{i_2}(z_2)\ldots A_{i_n}(z_n)\ket{0}$$
followed by taking successively the limits, $z_n\rightarrow 0$, then
$z_{n-1}\rightarrow 0$, etc. and finally extracting the leading term in that
limit. As an example, with this prescription (we consider the NS sector, only
for simplicity)
$$(\ppu\pmd)^3\ket{0}\equiv\ppu_{-5/2}(\pmd)_{-5/2}\ppu_{-3/2}(\pmd)_{-3/2}
\ppu_{-1/2}(\pmd)_{-1/2}\ket{0}.$$
This notation implies that we have defined the action of the operator,
$K(p)$ for all states.
It is easy to verify that the state $\ket{p}$ has the following properties
\bea
\ppu_r\ket{p}=(\pmd)_r\ket{p}=0,&&r\ge -3p+\hf\nn
(\ppd)_r\ket{p}=\pmu_r\ket{p}=0,&&r\ge +3p+\hf\nn
V^+_N\ket{p}=(V_-)_N\ket{p}=0,&&N\ge -3p\nn
(V_+)_N\ket{p}=V^-_N\ket{p}=0,&&N\ge +3p\nn
(V^+_{-3p-1})^n\ket{p}=[(V_-)_{-3p-1}]^n\ket{p}=0,&&n>\ktp .
\label{eq:ketp}
\eea
With this it is easy to prove that the states, $\ket{p}$ are all massless
$\At$ highest weight states. Further they are highest weight states of the
$U(1)$ algebra of $Y(z)$, and finally they have the quantum numbers
\bea
\lp &=&\lm =0\nn
Y&=&2pk=2u/\sqrt{3}\nn
h&=&3p^2k\nn
h^Y&=&\frac{u^2}{k}=3p^2k=h,
\eea
so that the conformal dimension $\tilde{h}$ pertaining to $\At$ is
$\tilde{h}=h-h^Y=0$ independent of $p$.
Hence, indeed these states furnish a representation of $\Ak$ with $m=0$.

It is simple to see that this technique in fact works also when $K(p)$ acts
on other $\At$ highest weight states: the resulting state is (i) also
an $\At$ highest weight state with the same $\At$ quantum numbers, (ii)
it is a highest weight state for $Y(z)$, and (iii) the set of all these for
$p\in{\sf Z}$ provides a representation of $\Ak$ with
$$m\equiv 3Y\ \mbox{mod} \ \ 6k$$
where $Y$ is the total ``hypercharge" of any of the states, $K(p)\ket{hws}$.

This technique implies that in the following we only have to understand one
$\At$ highest weight state for each set of $\At$ quantum numbers and $m$.
The operator $K(p)$ will furnish the infinite tower corresponding to the
representation of $\Ak$.

\subsection{Massless $\At$ states}
We now provide a complete list of all massless $\At$ highest weight states
in ${\cal H}(\ktp ,\Lambda)$. For $\va$ an $SU(3)$ weight corresponding to
an integrable representation (for the given integer value of $\ktp$),
i.e. $a_i$ non zero integers such that,
$$0\le a_1+a_2\le \ktp,$$
let,
$$\ket{\va ;p}=K(p)\ket{\va ;0},$$
the last state being,
$$\ket{\va ;0}\equiv \ket{\va}\otimes\ket{0},$$
the tensor product of the $\SSS$ highest weight state, $\ket{\va}$ and
the WS Fock space vacuum, $\ket{0}$. In fact, according to the above we only
need consider the $p=0$ case, the rest being trivially obtained by acting
with $K(p)$.

The complete list of massless states for
$$\Lambda = (\va ,\ktp ,0)$$
is as follows,
\bea
\ket{(1)\va ;p}&\equiv&
\ppd\pmu(V^-)^{\ktp-(a_1+a_2)}(V_+)^{a_1}\ket{\va ;p}\nn
\ket{(2)\va ;p}&\equiv&
\ppu\pmd (V^+)^{\ktp-(a_1+a_2)}(V_-)^{a_2}\ket{\va ;p}\nn
\ket{(3)\va ;p}&\equiv& \ket{\va ;p}\nn
\ket{(4)\va ;p}&\equiv&\pmd(V_-)^{a_2}\ket{\va ;p}\nn
\ket{(5)\va ;p}&\equiv&\ppd (V_+)^{a_1}\ket{\va ;p}\nn
\ket{(6)\va ;p}&\equiv&\pmd\ppu\pmd(V^+)^{\ktp-(a_1+a_2)}(V_-)^{a_1+a_2}
\ket{\va ;p}.
\label{eq:mlstates}
\eea
The quantum numbers of these states are as follows,
\bea
2\lm_i&=&0, \ \ i=1,2,3\nn
2\lm_i&=&1, \ \ i=4,5,6\nn
2\lp_1(\va )=\ktp-a_1,&&m_1(\va)=a_1-a_2-3(\ktp -a_2)\nn
2\lp_2(\va)=\ktp -a_2,&&m_2(\va )=a_1-a_2+3(\ktp -a_1)\nn
2\lp_3(\va )=a_1+a_2,&&m_3(\va )=a_1-a_2\nn
2\lp_{i+3}=\ktp -2\lp_i (\va ),&&m_{i+3}(\va )=m_i(\va) +3k, \ \ i=1,2,3.
\eea
The states considered in ref.~{\cite{gptvp}} were
$$\ket{(3)\va ;0},\ \ \ket{(4)\va ; 0},\ \ \ket{(5)\va ;0}.$$
These were enough to exhaust the quantum number possibilities for massless
$\At$
states, however here we must find all massless $\At$ states in
${\cal H}(\Lambda ,\ktp )$. That the above list is exhaustive is partly
conjectural. In this connection as well as in general, however, it is
interesting to notice a particularly relevant property of these states when
represented in terms of the standard fermionic ``quark model" of the $\SSS$
current algebra.

Thus consider $3\times \ktp$ complex fermion pairs,
$$q_i^f (z), \ \oq_i^f(z)$$
with $i=1,2,3$ being $SU(3)$ triplet indices and $f=1,2,\ldots ,\ktp$ being
``flavour" indices. Our $\SSS$ representation spaces may be thought of as
lying in the subspace of the Fock space of the quarks which is a  flavour
singlet. We shall understand the relevant projection to flavour singlets
being performed whenever necessary. By $\ket{0}$ we now understand the
combined
vacuum of the quarks and the WS fermions. The state, $\ket{\va ;0}$ may
then be thought of as the following (or rather a combination of such states
to make a flavour singlet)
$$\ket{\va ;0}= (q_1^{f_1})_{-\hf}\cdots (q_1^{f_{a_1}})_{-\hf}
(\oq_2^{f'_1})_{-\hf}\cdots (\oq_2^{f'_{a_2}})_{-\hf}\ket{0}$$
the sets, $\{f_1,\ldots ,f_{a_1}\}$ and $\{f'_1,\ldots ,f'_{a_2}\}$ being
non-overlapping.
Similarly the $\SSS$ generators entering the supercurrents are
\bea
V^+\sim\oq_3^f q_1^f&&V_+\sim \oq_1^f q_3^f\nn
V^-\sim \oq_2^f q_3^f && V_-\sim \oq _3^f q_2^f
\eea
sum over $f$ implied.

The states, $\ket{p}$ above then have the following further properties
\bea
(q_1^f)_r\ket{p}=(q_2^f)_r\ket{p}=0,&& r\ge -p+\hf,\nn
(\oq_1^f)_r\ket{p}=(\oq_2^f)_r\ket{p}=0,&& r\ge +p+\hf,\nn
(\oq_3^f)_r\ket{p}=0,&&r\ge -2p+\hf,\nn
(q_3^f)_r\ket{p}=0,&&r\ge +2p+\hf.
\eea
These relations together with those in \eqn{ketp} show that the states
$\ket{p}$ have the nice property of simply representing shifted vacuum levels
for all the fermions -- WS ones and quarks alike --  evenly in all flavours.
Likewise, all the massless states given in \eqn{mlstates} have the property
that they represent shifted fermionic vacuum states, however with the vacuum
levels shifted by amounts depending on flavour. More precisely, linear
combinations of such states corresponding to projections to flavours singlets.
Finally we point out that the enumerations of the massless $\At$ highest weight
states in \eqn{mlstates} immediately allow us to write down the expression
for the matrix, ${M_{\LA}}^L(q,z_y)$, \eqn{matrix}, by inspection.

\subsection{Massive $\At$ states}

{}From the above discussion several points concerning the embedding of massive
highest weight $\At$ states in ${\cal H}(\Lambda ,\ktp)$ immediately emerge.

First our general argument demonstrates that for given $\At$ quantum numbers,
they too come in infinite multiplets corresponding to a representation of
$\Ak$. Second, our studies of modular properties of characters as summarized
in Section 4, imply that for given $\lp$ ($\lm $ is necessarily $0$ for massive
representations in the NS sector and $0\le 2\lp \le \ktp-1$, cf.
ref.~{\cite{gptvp}}), there must be a denumerable infinity of massive $\At$
highest weight states corresponding to a certain infinite set of $\tilde{h}$
values. This follows because we know that the characters of $WS\otimes\SSS$
transform according to a finite representation of the modular group whereas
that will be possible only if infinitely many $\At$ characters are involved on
the right hand side of the sumrule for $\{\chi^{WS,NS}\cdot \chi_{\LA}\}_m$
in eq. \eqn{massivesum}. It is completely outside the scope of the present
paper to provide an understanding as to how these infinitely many massive
$\At$ representations are embedded in detail in $\HLk$. It is interesting
to remark, however, that in the quark model description introduced above
they will all correspond to certain (solitonic) excitations above the shifted
vacua corresponding to the massless states.

Also, for a given $\SSS$ representation $\Lambda$ containing the $SU(3)$
representation, labelled by $\va$, and for given $\lp$ it is not too hard
to provide the rule to work out what $m$ values will occur in the sum rule
as far as the representations of $\Ak$ are concerned. To this we now turn.

First notice that if we have imbedded in $\HLk$ a representation of
$\At\oplus\Ak$ with quantum numbers $\lp$ for $\At$ ($\lm=0$) and
$\overline{m}$ for $\Ak$,
then we can expect all the ones with the same $\lp$ and with
$m =\overline{m}+6\Zi_k$ to arise as well.
The corresponding states may be obtained from
the highest weight of the first one by a suitable operator, and since neither
$\lp$ nor $\lm$ is allowed to change that operator
must involve an integer number of
{\em pairs} $(V^+,V_-)$ and $\ppu\pmd$ (or their conjugates), each pair
contributing an even increment to $Y$ and thus a multiple of 6 to $m$.

To determine the offset in $m$ away from $6\Zi_k$, we notice that the highest
weight state in the representation of $\At\otimes\Ak$ must have $(\lp ,Y)$
quantum numbers lying in the weight lattice of the $SU(3)$ representation in
question. A little consideration shows that this implies that the rule
for obtaining the possible $m$ values for given $\Lambda$ and $\lp$ is as
follows.
Let
$$t\equiv a_1-a_2\ \ \mbox{mod}\ \ 3$$
be the triality of the $SU(3)$ representation, then
\ben
m=\left\{
\begin{array}{ll}
-2t+6\Zi_k&\mbox{for} \ 2\lp \ \mbox{even}\\
3-2t+6\Zi_k&\mbox{for} \ 2\lp \ \mbox{odd}.
\end{array}\right.
\een
This means that we may write the massive part of the sumrule,
\eqn{massivesum} as follows,
\bea
&&\{\chi^{WS,NS}(q,z_-,z_y)\chi_{\LA}(q,z_+,z_y)\}_m\nn
&=&~~~\sum_{2\lp =0}^{\ktp -1}Ch^{{\At},NS}_m(\lp ,q,z_+,z_-)
\sum_{n\in\Zi_k}\chi^{3k}_{-2(a_1-a_2)+6n+6\lp}(q,z_y)F^{\LA}_{2\lp ,n}(q),
\eea
where the massive $\At$ character is taken at the special value of conformal
dimension $\tilde{h}_m$ of the previous section, and where the set of massive
representations arising is given by the expansion of the modular forms,
$F^{\LA}_{2\lp ,n}(q) $ in  a power series in $q$ with positive integer
coefficients, apart from some overall prefactor involving a fractional
power of $q$.

This justifies \eqn{matrixt}.

\section{Modular Properties of $\At$ characters}
\setcounter{equation}{0}

The very form of the massive $\At$ characters \eqn{nsmassive} allows for a
straightforward derivation of their modular properties in terms of the
modular transformations of $\widehat{SU(2)}_{\ktp }\times
\widehat{SU(2)}_1$ and Virasoro
characters at level $\ktp +1$. Because the combinations of massless $\At$
characters $Ch_0^{\At,NS}(L)$ do not have the simple form of the massive ones,
their modular transformations are not so easily obtained. If one assumes that
the massive and massless sectors of the sumrules \eqn{charsum} decouple from
each other under S and T, the number of equations for the determination of the
unknown modular transformation matrices by far exceeds the number of unknowns,
more and more so as $\ktp\rightarrow\infty$.
Thus, finding a consistent solution
we take as a strong indication for the validity of the conjecture.
A consistent solution
is one compatible with the modular transformations of the LHS of the
sumrules, which involve $\SSS$. We will show below that it implies
the remarkable result, that the combinations
of massless characters transform like $\widehat{SU(2)}_{\ktp +1}$ characters.
Indeed, as was pointed out in \cite{opt}, when the angular variables $z_\pm$
are suitably correlated, they do indeed reduce to such characters, but for
general values of $z_\pm$ that is not the case.
Furthermore,
the decoupling hypothesis implies that the functions
$F^{\LA }_{2\lp ,n}(q)$ introduced in \eqn{massivesum}
carry a finite representation of the modular group. When multiplied by
$1/\eta$,
these functions have the right properties to be the characters (or linear
combinations thereof) of a rational
conformal field theory with central charge
\ben
c_{\phi }=1+3(\ktp-1)^2/(\ktp +1)(\ktp +3).
\een
In the case $\ktp =1$, as explained in Appendix A, the functions
$F^{\LA }_{2\lp,n}(q)$ are completely determined. The decoupling may be
verified and the combinations of massless characters actually transform as
$\widehat{SU(2)}_2$ characters. We leave for further investigation \cite{next}
the general
formula for the modular transformations of the functions $F^{\LA }_{2\lp,n}(q)$
and only give here the result for the next theory in the series ($\ktp =2$).
If the functions $F^{\LA }_{2\lp ,n}(q)$ are organized in a column vector
\ben
\frac{1}{\eta }(F^{((0,0),2,0)}_{0,0},F^{((0,0),2,0)}_{0,1},
F^{((0,0),2,0)}_{0,2},
F^{((1,1),2,0)}_{0,0},F^{((1,1),2,0)}_{0,1},F^{((1,1),2,0)}_{0,2})^T,
\een
the matrix of their S transform is
\[ S = \frac{2}{5}\left[ \begin{array}{cccccc}
 s_3 & 2s_3    & 2s_3      & s_6  & 2s_6     & 2s_6     \\
 s_3 & 2s_3c_6 & -s_6      & s_6  & s_{12}   & -2s_6c_3 \\
 s_3 & -s_6    & 2s_3c_6   & s_6  & -2s_6c_3 & s_{12}   \\
 s_6 & 2s_6    & 2s_6      & -s_3 & -2s_3    & -2s_3    \\
 s_6 & s_{12}  & -2s_6c_3  & -s_3 & -2s_3c_6 & s_6      \\
 s_6 &-2s_6c_3 & s_{12}    & -s_3 & s_6      & -2s_3c_6
\end{array} \right],   \]
where we have defined $s_n=\sin \frac{n\pi }{15}$ and
$c_n=\cos \frac{n\pi }{15}$. The above matrix is not orthogonal such as
would be expected for the transformation matrix for a set of characters.
However, it seems very likely that these modular functions are instead
combinations of characters of some extended algebra, with the reduction of
angular variables implying symmetries that have forced us to look at such
combinations only. Indeed in the limit $z_y=1$ the characters of the rational
torus model $\Ak$ exhibit this very behaviour.

Let us now concentrate on the massless sector and show how the
$\widehat{SU(2)}_{\ktp+1}$
transformation laws emerge as a consistent solution to the modular
transformations of the sumrules.
The S transform of the massless sector of the sumrule \eqn{masslesssum}
for a weight $\vec{a}$ of triality $t$ is given by (we set $z_y=1$ in the
following for simplicity)
\ben
\{ \chi^{WS,NS}\cdot \sum_{\epsilon =0,\pm 1} \sum_{\vec{a}' \in (\epsilon =0)~
{\rm region} }
[S(t,\epsilon )]_{\vec{a}}^{\phi ^{\epsilon }(\vec{a}')}
{}~\chi_{\phi ^{\epsilon}(\LA')}(\epsilon )\}_0
=(M(t)^S)_{\vec{a}}^L~(Ch_0^{\At,NS})^S(L),
\label{eq:Stransf}
\een
where we use the S transform of the affine $\SSS$ characters in the
form
\ben
(\chi_{\LA})^S = S_{\vec{a}}^{\vec{a}'}~ \chi_{\LA'},
\een
with $\LA '=(\vec{a}',\ktp,0)$ and
$$\phi^{\epsilon}(\LA)\equiv (\phi^{\epsilon}(\va) ,\ktp ,0).$$
Further,
\ben
S_{\vec{a}}^{\vec{a}'}=\Sigma _{\vec{a}}^{\vec{a}'}+
(\Sigma_{\vec{a}}^{\vec{a}'_C})^*,
\label{eq:smatrix1}
\een
and
\ben
\Sigma_{\vec{a}}^{\vec{a}'}=\frac{i}{k\sqrt 3}
z_{3k}^{2(\alpha _1-\alpha_2)(\alpha _1'-\alpha_2')}
(z_k^{2\alpha_1\alpha_2'}-1)(z_k^{2\alpha_2\alpha_1'}-1).
\label{eq:smatrix2}
\een
The notations are $z_k=e^{\frac{i\pi }{k}}$,
$\alpha_i=a_i+1$, and $\vec{a}_C=(a_2,a_1)$ when
$\vec{a}=(a_1,a_2)$. Also we write $\LA_C\equiv(\va _C,\ktp ,0)$.

In the above formula, the weights $\vec{a}'$ correspond to integrable $\SSS$
representations and are classified according to
the $\epsilon $ region to which they belong. By definition, a weight $\vec{a}'=
(a'_1,a'_2)$ belongs to the region $\epsilon =0$ if it is the representative
of its orbit under the transformation $\phi $ \eqn{phi} with the lowest
$a'_1+a'_2$ value. If two weights within the same orbit have the same
$a'_1+a'_2$ value, then the weight with the highest $a'_1$ value will
belong to the $\epsilon =0$ region. One naturally obtains weights
in the $\epsilon = \pm 1$ regions by applying $\phi ^{\pm 1}$ to the
$\epsilon =0$ weights.
{}For a weight $\vec{a}$ of triality $t$, and $\vec{a}'$ having a value of
$\epsilon $, one may write
\ben
S(t,\epsilon)_{\vec{a}}^{\vec{a}'}=y^{-t\epsilon}S(t)_{\vec{a}}^{\vec{a}'},
\een
where $y=e^{\frac{2i\pi }{3}}$. The unitarity condition then takes the form
\ben
\sum_{\epsilon}[\sum_{\vec{a}'}S(t)_{\vec{a}}^{*~\vec{a}'}
S(t')_{\vec{a}}^{\vec{a}'}]y^{\epsilon (t'-t)}=1.
\een
It is clear that the above expression vanishes whenever $t \ne t'$, whereas
for $t=t'$, the sum over $\epsilon $ merely gives a factor 3. It follows
that the {\em reduced} matrices $S(t)$ are normalized so that
$S(t)S(t)^{\dagger}=1/3$ for all $t$. A complication occurs whenever
$\ktp$ is a multiple of 3 and $t \ne 0$ since the matrix $S(t)$ is not square,
having an extra column corresponding to the central representation with
Dynkin labels $(\frac{\ktp}{3},\frac{\ktp}{3})$. One however can prove that
the matrix element of $S(t \ne 0)$ having this representative as column
label vanishes.

If one defines the S transform of combinations of massless $\At$ characters as
\ben
(Ch_0^{\At,NS}(L))^S=S_L^{L'}~Ch_0^{\At,NS}(L'),
\een
the above considerations allow to rewrite the transformation law \eqn{Stransf}
as the following ``master equation "
\ben
3\sum_{\vec{a}}(S(t)_{\vec{a}}^{~\vec{a}'})^*(M(t)^S)_{\LA}^{~L}=
\sum_{\epsilon ,L'}M_{\phi ^{\epsilon }(\LA ')}^{~L'}
(\epsilon )~y^{-t\epsilon}
S_{L'}^{~L}.\label{eq:master}
\een
Here, the modular transformation matrix for the $SU(3)$ characters is the
reduced one pertaining to $\epsilon =0$. The index $\vec{a}$ runs over
triality $t$ representations only, whereas $\vec{a}'$ is a representation
pertaining to $\epsilon =0$. For $\ktp $ a multiple of 3 and $t \ne 0$,
it is understood that the representation $\vec{a}'=(\frac{\ktp}{3},
\frac{\ktp}{3})$ is removed.
As explicitly shown in Appendix C, a solution to the above equation is
provided by
\ben
S_{L'}^{~L}=-\frac{2}{k}\sin \left(\frac{\pi}{k}(L+1)(L'+1)\right),
{}~~~L,L'=0,...,k-2,
\een
which corresponds, up to a sign, to the elements of the S transform matrix
for $\widehat{SU(2)}_{\ktp+1}$ characters.\\[1.0cm]

\section{Conclusions}
In a series of previous publications \cite{pt1,pt2,pt3,opt} we provided
a rather detailed study of the characters for the unitary highest
weight representations of the $N=4$ doubly extended superconformal algebra
$\At$. In this paper we have presented a number of investigations
aimed at understanding field theories based on this algebra. To this end
we have made use of previously known coset constructions
realizing the algebra $\At$
\cite{stvp88,vp,gptvp,theo}. We have concentrated on the
family based on the Wolf space $SU(3)/SU(2)\times U(1)$ which
corresponds to one
of the two central extensions being unity ($\ktm$ in our case) but the other,
$\ktp$, being arbitrary. We have studied the character sumrules induced
by this realization, and our main results are as follows.

First it
has been possible to obtain information on the modular transformation
properties of the massless characters. Remarkably,
certain combinations of massless characters transform exactly as $SU(2)$
characters of level $\ktp +\ktm$. This result was mentioned in \cite{opt}
where it emerges from a completely different point of view, and
some of the theorems stated there are proven in this paper.

Second, we show in two different ways that the realization
gives rise to $\At\times U(1)$, or more interestingly to the particular
rational extension $\Ak$ of $U(1)$ having an extra dimension $3k=3(k^++k^-)$
operator in the algebra \cite{dijk}. This result is
analogous to what happens when one studies the branching of affine $SU(3)$
characters into $SU(2)$ ones \cite{Huitu}, and in fact provides
one way of proving that result. Perhaps more importantly this reduction
allows the derivation of the modular properties of the
combinations of massless characters.

Third, the massive part of the sumrule is shown to be written as
infinite sums of massive $\At$ characters, which not only
involve the $\Ak$ characters again, but also certain modular
forms $F^{\LA}_n$, which appear to transform according to a finite
dimensional representation of the modular group. These forms
multiply the massive
characters evaluated at a particular conformal dimension, which is
chosen in such a way that the massive characters
also transform according to a finite dimensional representation of the modular
group. This strongly implies that these modular forms are intimately
related to a separate rational conformal field theory, a suggestion which is
further strengthened by comparison with a recent free field realization of
these algebras \cite{imp}. We have presented several general properties of
these modular forms and we have demonstrated their relation to a variety
of branching functions. These studies have revealed a number of remarkable
identities. In special cases it has been possible to relate some of
the $F^{\LA}_n$ to other known modular forms.
It appears that they
are also related to coset theories based on certain of
the Kazama-Suzuki $N=2$ theories \cite{KS}.
This last fact is potentially significant and
deserves much further investigation.

In conclusion, we have made progress towards understanding the
modular properties of the $\At$ characters in general
and towards an
understanding of the rational conformal field theories involved in connection
with certain coset realizations in particular. This latter subject
contains a rich set of interesting structures relating the conformal field
theories with a chiral $\At$ algebra to several
other known but a priori unrelated conformal field theories.

\newpage
\noindent {\bf Acknowledgements}\\[0.3cm]
We would like to thank P. Bowcock, E.F. Corrigan, A. Kent, W. Lerche,
J.O. Madsen, H. Ooguri,
F. Ravanini and G. Watts for stimulating discussions, and particularly
F.A. Bais and T. Eguchi for their insight in coset models and the
Kazama-Susuki construction. We also thank the Aspen Centre for Particle
Physics where this work was started. AT acknowledges the Science
and Engineering Research Council for an Advanced Fellowship. JLP acknowledges
support from EEC contract SC1 394 EDB.

\vskip 2cm
\appendix
\section{Determination of the functions $P_{2\lp ,n }^{\LA}(q)$ and
$F_{2\lp ,n}^{\LA}(q)$ for $\ktp =1,2$ and $\ktm =1$}

\setcounter{equation}{0}
\subsection{The case $\ktp =1$}

When $\ktm =\ktp =1$, i.e. $k=4$, the three integrable
$\widehat{SU(3)}_1$ representations to consider pertain to the single
orbit of the transformation $\phi $ \eqn{phi}. It is therefore sufficient
to consider the sumrule \eqn{charsum} for one of them (say the singlet)
to determine all functions $P_{2\lp ,n}^{\LA}(q)$ and
$F_{2\lp ,n}^{\LA}(q)$. The symmetries \eqn{Fsym1}, \eqn{Fsym2},
\eqn{Fsym3} and \eqn{Psym}
described in Section 2 allow for a particularly simple sumrule,
\bea
&&\sum_{2\lm =0}^{1}\sum_{2\lp =0}^{1} \chi_{2\lp }^{1}(q,z_+)
\chi_{2\lm }^{1}(q,z_-)
P_{0,0 }^{((0,0),1,0)}(q)
 \sum_{r=0}^3
\frac{1}{\eta}\theta_{2(\lm +r)-6\lp ,4}(q)\chi^{12}_{6(r+\lm +\lp )}(q,z_y)
=\nn
&-&Ch_0^{\At,NS}(0,q,z_+,z_-)[\chi_3^{12}(q,z_y)+\chi_{-3}^{12}(q,z_y)]\nn
&+&Ch_0^{\At,NS}(1,q,z_+,z_-)[\chi_0^{12}(q,z_y)-\chi_{12}^{12}(q,z_y)]\nn
&+&Ch_0^{\At,NS}(2,q,z_+,z_-)[\chi_9^{12}(q,z_y)+\chi_{-9}^{12}(q,z_y)]\\
&+&Ch_m^{\At,NS}(0;q,z_+,z_-)
\left\{F_{0,0}^{((0,0),1,0)}(q)[\chi_0^{12}(q,z_y) +\chi_{12}^{12}(q,z_y)]
\right.\nn
      &+&\left. F_{0,1}^{((0,0),1,0)}(q)[\chi_6^{12}(q,z_y)
      +\chi_{-6}^{12}(q,z_y)] \right\}.\nonumber
\eea
In order to determine the functions $P_{0,0}^{((0,0),1,0)}(q)$
and $F_{0,n}^{((0,0),1,0)}(q), n=0,1$, it is sufficient to identify the
coefficients of the rational torus characters
$\chi_0^{12}(q,z_y)$ and
$\chi_{6}^{12}(q,z_y)$  on both sides of the sumrule.
Recalling that
$\theta _{m,k}(q)=\theta _{-m,k}(q)=\theta_{2k-m,k}(q)$,
one gets
\bea
\lefteqn{Ch_0^{\At,NS}(1,q,z_+,z_-)
+F_{0,0}^{((0,0),1,0)}(q)Ch_m^{\At,NS}(0,q,z_+,z_-)}\nn
&=& \frac{1}{\eta}
P_{0,0}^{((0,0),1,0)}(q)[\theta_{0,4}(q)\chi_0^1(q,z_+)\chi_0^1(q,z_-)
                    + \theta_{4,4}(q)\chi_1^1(q,z_+)\chi_1^1(q,z_-)],
\eea
and
\ben
F_{0,1}^{((0,0),1,0)}(q)Ch_m^{\At,NS}(0,q,z_+,z_-)=\frac{1}{\eta}
\theta_{2,4}(q) P_{0,0}^{((0,0),1,0)}(q)[\chi_0^1(q,z_+)\chi_0^1(q,z_-)
                                         +\chi_1^1(q,z_+)\chi_1^1(q,z_-)].
\een
Next, one can use the decomposition of $\At$ characters in
$SU(2)_1 \times SU(2)_1$
characters \eqn{nsmassive},\eqn{branching}, namely
\ben
Ch_m^{\At,NS}(0,q,z_+,z_-)=\frac{1}{\eta }[\chi_0^1(q,z_+)\chi_0^1(q,z_-)
+\chi_1^1(q,z_+)\chi_1^1(q,z_-)],
\een
\ben
Ch_0^{\At,NS}(1,q,z_+,z_-)=Y_{1/16}^{(3)}(q)[\chi_0^1(q,z_+)\chi_0^1(q,z_-)
-\chi_1^1(q,z_+)\chi_1^1(q,z_-)],
\een
where $Y_{1/16}^{(3)}(q) \equiv Y_{1,2}^{(3)}(q)=Y_{2,2}^{(3)}(q)$.
By identifying
the coefficients of the $SU(2)_1$ bilinears in the relations above,
one obtains the following results,
\bea
F_{0,0}^{((0,0),1,0)}(q)&=&
\hf [\theta_{0,4}(q)+\theta_{4,4}(q)] P_{0,0}^{((0,0),1,0)}(q)\nn
F_{0,1}^{((0,0),1,0)}(q)&=&
\theta_{2,4}(q) P_{0,0}^{((0,0),1,0)}(q)\nn
Y_{1/16}^{(3)}(q)&=&\frac{1}{2\eta} P_{0,0}^{((0,0),1,0)}(q)
[\theta_{0,4}(q)-\theta_{4,4}(q)].
\eea
One can see how the functions $P_{0,0 }^{((0,0),1,0)}(q)$ and
$F_{0,n}^{(\va ,1,0)}(q)$
are related to the branching function $Y_{1/16}^{(3)}(q)$. The nice feature
of the
particular case $\ktm =\ktp =1$ is that it is possible to completely determine
the branching functions $Y_{r,s}^{(3)}(q)$, as we shall now see.
It was noticed in \cite{opt} that the
Ramond massless $\At$ characters reduce to $\SU$ characters when
$z_-=-z_+^{-1}$, a property which, in the NS sector and when $\ktm =\ktp =1$,
can be expressed as
\ben
Ch_0^{\At,NS}(L,q,z_+,z_-=-q^{-\hf }z_+^{-1})=-q^{-1/4}z_+^{-1}\chi^2_L(q,z_+).
\label{eq:su2}\een
The decomposition formula \eqn{branching} when $z_-=-q^{-\hf }z_+^{-1}$
gives the following set of relations,
\bea
\chi^2_0(q,z_+)&=&Y_0^{(3)}\chi_0^1(q,z_+)\chi^1_0(q,z_+)+
                Y_{\hf }^{(3)}\chi_1^1(q,z_+)\chi_1^1(q,z_+),\nn
\chi^2_1(q,z_+)&=&2Y_{1/16}^{(3)}\chi_0^1(q,z_+)\chi^1_1(q,z_+),\nn
\chi^2_2(q,z_+)&=&Y_{\hf }^{(3)}\chi_0^1(q,z_+)\chi^1_0(q,z_+)+
                Y_0^{(3)}\chi_1^1(q,z_+)\chi_1^1(q,z_+),
\label{eq:relations}
\eea
where we have used the well-known fact,
\ben
\chi_{2\ell }^{k}(q,-q^{-\hf }z^{-1})=(-1)^{2\ell }q^{-1/4}z^{-1}
\chi_{k-2\ell }^k(q,z),
\een
and introduced the notation
$Y_0^{(3)}(q)\equiv Y_{1,1}^{(3)}(q)=Y_{2,3}^{(3)}(q)$
and $Y_{\hf }^{(3)}(q)\equiv Y_{2,1}^{(3)}(q)=Y_{1,3}^{(3)}(q)$.
The GKO character sumrules \cite{GKO} allow to solve for the branching rules
with the result,
\bea
Y_0^{(3)}(q)&=&\frac{[\chi_0^{Vir,(3)}]}
{[\chi_0^{Vir,(3)}]^2-[\chi_{\hf }^{Vir,(3)}]^2}=\eta ^{-1}\theta_{1,4}(q),\nn
Y_{\hf }^{(3)}(q)&=&-\frac{\chi_{\hf }^{Vir,(3)}}
{[\chi_0^{Vir,(3)}]^2-[\chi_{\hf }^{Vir,(3)}]^2}=-\eta ^{-1}\theta_{3,4}(q),\nn
Y_{1/16}^{(3)}(q)&=&\frac{1}{2\chi_{1/16}^{Vir,(3)}}
=\hf \eta ^{-1}[\theta_{0,4}(q)-\theta_{4,4}(q)].
\eea
It is now straightforward to obtain the analytic results summarized in the
following,
\bea
P_{0,0}^{((0,0),1,0)}(q)&=&P_{1,0}^{((0,0),1,0)}(q)=1,\nn
F_{0,0}^{((0,0),1,0)}(q)&=&F_{0,2}^{((0,0),1,0)}(q)
=\hf [\theta _{0,4}(q)+\theta _{4,4}(q)]=\frac{\eta }{2}\chi^1_0(q),\nn
F_{0,1}^{((0,0),1,0)}(q)&=&F_{0,3}^{((0,0),1,0)}(q) =
\theta _{2,4}(q)=\frac{\eta }{2}
\chi^1_1(q).
\eea
The functions corresponding to the non singlet representations are obtained
by implementing the symmetries \eqn{Fsym3}, \eqn{Psym}.
The results above
were presented in a different form in \cite{opt}. In this new
derivation, it is remarkable that the functions $F_{0,n}^{(\va ,1,0)}(q)$
and the
branching functions $Y_{r,s}^{(3)}(q)$ span a basis for theta functions at
level 4. The occurrence of the latter is obviously deeply rooted in the
relation \eqn{identity} which links theta functions at level 1 and 3 to
theta functions at level 4 and 12. As we shall see when $\ktp =2$,
the theta functions at level 10 enter in the expressions for
$F_{2\lp ,n}^{(\va ,2,0)}$
and $Y_{m,n}^{(4)}$, but in a much more complicated pattern.

\subsection{The case $\ktp =2$}

The six integrable $\widehat{SU(3)}_2$
representations organize themselves in two orbits
of the transformation $\phi $, and it is sufficient to consider the sumrules
corresponding to one representative of each orbit, say the singlet and the
octet, for which the sumrules take the following form,
\bea
&&\sum_{2\lm =0}^1\sum_{2\lp =0}^2
\chi^2_{2\lp }(q,z_+)\chi^1_{2\lm }(q,z_-)\nn
&\times&\sum_{n=0}^1 P_{2\lp ,n}^{((0,0),2,0)}(q)
\sum_{r=0}^4\frac{1}{\eta}\theta_{4(\lm +r)-6(n+\lp ),10}(q)
\chi^{15}_{6(\lp +\lm +n+r)}(q,z_y)
=\nn
&-&Ch_0^{\At,NS}(0,q,z_+,z_-)[\chi_3^{15}(q,z_y)+\chi_{-3}^{15}(q,z_y)]
+Ch_0^{\At,NS}(1,q,z_+,z_-)\chi_0^{15}(q,z_y)\nn
&-&Ch_0^{\At,NS}(2,q,z_+,z_-)\chi_{15}^{15}(q,z_y)
+Ch_0^{\At,NS}(3,q,z_+,z_-)[\chi_{12}^{15}(q,z_y)+\chi_{-12}^{15}(q,z_y)]\nn
&+&Ch_m^{\At,NS}(0;q,z_+,z_-)\nn
&\times&\left\{
F_{0,0}^{((0,0),2,0)}(q)\chi_0^{15}(q,z_y)+F_{0,1}^{((0,0),2,0)}(q)
[\chi_6^{15}(q,z_y)+\chi_{-6}^{15}(q,z_y)]\right.\nn
&+&\left.
F_{0,2}^{((0,0),2,0)}(q)[\chi_{12}^{15}(q,z_y)+\chi_{-12}^{15}(q,z_y)]\right \}
\nn
&+&Ch_m^{\At,NS}\left(\hf ;q,z_+,z_-\right)\nn
&\times&\left \{
F_{0,0}^{((0,0),2,0)}(q)\chi_{15}^{15}(q,z_y)
   +F_{0,1}^{((0,0),2,0)}(q)[\chi_9^{15}(q,z_y)
+\chi_{-9}^{15}(q,z_y)]\right.\nn
&+&\left. F_{0,2}^{((0,0),2,0)}(q)
[\chi_3^{15}(q,z_y)+\chi_{-3}^{15}(q,z_y)]\right \},\nn
\eea
and
\bea
&&\sum_{2\lm =0}^1\sum_{2\lp =0}^2 \chi^2_{2\lp }(q,z_+)
\chi^1_{2\lm }(q,z_-)\nn
&\times&\sum_{n=0}^1 P_{2\lp ,n}^{((1,1),2,0)}(q)
\sum_{r=0}^4\frac{1}{\eta}\theta_{4(\lm +r)-6(n+\lp ),10}(q)
\chi^{15}_{6(\lp +\lm +n+r)}(q,z_y)
 =\nn
&-&Ch_0^{\At,NS}(0,q,z_+,z_-)\chi_{15}^{15}(q,z_y)
-Ch_0^{\At,NS}(1,q,z_+,z_-)[\chi_6^{15}(q,z_y)+\chi_{-6}^{15}(q,z_y)]\nn
&+&Ch_0^{\At,NS}(2,q,z_+,z_-)[\chi_{9}^{15}(q,z_y)+\chi_{-9}^{15}(q,z_y)]
+Ch_0^{\At,NS}(3,q,z_+,z_-)\chi_0^{15}(q,z_y)\nn
&+&
Ch_m^{\At,NS}(0;q,z_+,z_-)\nn
&\times&\left\{
F_{0,0}^{((1,1),2,0)}(q)\chi_0^{15}(q,z_y)+F_{0,1}^{((1,1),2,0)}
[\chi_{6}^{15}(q,z_y)+\chi_{-6}^{15}(q,z_y)]\right.\nn
&+&\left. F_{0,2}^{((1,1),2,0)}(q)
[\chi_{12}^{15}(q,z_y)+\chi_{-12}^{15}(q,z_y)]\right \}\nn
&+&
Ch_m^{\At,NS}\left(\hf ;q,z_+,z_-\right)\nn
&\times&\left\{
F_{0,0}^{((1,1),2,0)}(q)\chi_{15}^{15}(q,z_y)
   +F_{0,1}^{((1,1),2,0)}(q)[\chi_9^{15}(q,z_y)
+\chi_{-9}^{15}(q,z_y)]\right.\nn
&+&\left. F_{0,2}^{((1,1),2,0)}(q)
[\chi_3^{15}(q,z_y)+\chi_{-3}^{15}(q,z_y)]\right \}.\nn
\eea
In the above, we have used the symmetries of the modular forms
$F_{2\lp ,n}^{(\va ,2,0)}$ \eqn{Fsym1}, \eqn{Fsym2}, \eqn{Fsym3}.
In order to determine
the parafermionic characters and to express $F_{2\lp ,n}^{(\va ,2,0)}(q)$
in terms  of the
branching functions $Y_{r,s}^{(4)}(q)$, it is sufficient to identify
in both sides of the sumrules the
coefficients of $\chi_0^2(q,z_+)\chi_0^1(q,z_-)$,
$\chi_1^2(q,z_+)\chi_1^1(q,z_-)$
and $\chi_2^2(q,z_+)\chi_0^1(q,z_-)$. To this end, we write, according to
\eqn{nsmassive},\eqn{branching},
\bea
Ch_m^{\At,NS}(0;q,z_+,z_-)&=&\eta ^{-1}[\chi_0^{Vir,(3)}(q)\chi_0^2(q,z_+)
\chi_0^1(q,z_-)+\chi^{Vir,(3)}_{1/16}(q)\chi_1^2(q,z_+)\chi_1^1(q,z_-) \nn
&+&\chi^{Vir,(3)}_{1/2}(q)\chi_2^2(q,z_+)\chi_0^1(q,z_-)],\nn
Ch_0^{\At,NS}(1,q,z_+,z_-)&=&Y^{(4)}_{1/10}(q)\chi_0^2(q,z_+)\chi_0^1(q,z_-)
-Y^{(4)}_{3/80}(q)\chi_1^2(q,z_+)\chi_1^1(q,z_-)\nn
&+&Y^{(4)}_{3/5}(q)\chi_2^2(q,z_+)
\chi_0^1(q,z_-),\nn
Ch_0^{\At,NS}(3,q,z_+,z_-)&=&Y^{(4)}_{3/2}(q)\chi_0^2(q,z_+)\chi_0^1(q,z_-)
-Y^{(4)}_{7/16}(q)\chi_1^2(q,z_+)\chi_1^1(q,z_-)\nn
&+&Y^{(4)}_{0}(q)\chi_2^2(q,z_+)
\chi_0^1(q,z_-),
\eea
where we introduced the notations
$Y^{(4)}_0\equiv Y^{(4)}_{1,1}=Y^{(4)}_{3,4}$,
$Y^{(4)}_{1/10}\equiv Y^{(4)}_{1,2}=Y^{(4)}_{3,3}$,
$Y^{(4)}_{3/5}\equiv Y^{(4)}_{1,3}=Y^{(4)}_{3,2}$,
$Y^{(4)}_{3/2}\equiv Y^{(4)}_{1,4}=Y^{(4)}_{3,1}$,
$Y^{(4)}_{7/16}\equiv Y^{(4)}_{2,1}=Y^{(4)}_{2,4}$ and
$Y^{(4)}_{3/80}\equiv Y^{(4)}_{2,2}=Y^{(4)}_{2,3}$.
Using exactly the same arguments as in the case $\ktp =1$,
one
can derive some relations among the branching functions $Y^{(4)}_{r,s}(q)$
(recall that in the previous case, the property \eqn{su2}
combined with the GKO
sumrules allowed for a complete determination of $Y^{(3)}_0,Y^{(3)}_{1/16}$
and $Y^{(3)}_{1/2}$),
\bea
Y^{(4)}_{7/16}&=&\frac{1}{\chi^{Vir,(3)}_{1/2}}
  [\chi^{Vir,(4)}_{1/10}-\chi^{Vir,(3)}_{1/16}Y^{(4)}_0],\nn
Y^{(4)}_{3/2}&=&-\frac{1}{\chi^{Vir,(3)}_{1/2}}
  [\chi^{Vir,(4)}_{3/80}-\chi^{Vir,(3)}_0Y^{(4)}_0],\nn
Y^{(4)}_{3/80}&=&-\frac{1}{\chi^{Vir,(3)}_{1/2}}
  [\chi^{Vir,(4)}_{3/2}+\chi^{Vir,(3)}_{1/16}Y^{(4)}_{3/5}],\nn
Y^{(4)}_{1/10}&=&\frac{1}{\chi^{Vir,(3)}_{1/2}}
  [\chi^{Vir,(4)}_{7/16}+\chi^{Vir,(3)}_0Y^{(4)}_{3/5}].
  \eea
A tedious but tractable computation provides us with an expression for the
parafermionic characters in terms of standard modular functions,
\bea
P_{0,1}^{((0,0),2,0)}&=&
\chi_{3/80}^{Vir,(4)}[\chi_{\hf }^{Vir,(3)}\theta _{4,10}
                                -\chi_0^{Vir,(3)}\theta_{6,10}]\Delta ^{-1},\nn
P_{0,0}^{((0,0),2,0)}&=&
\chi_{3/80}^{Vir,(4)}[\chi_0^{Vir,(3)}\theta _{4,10}
                           -\chi_{\hf }^{Vir,(3)}\theta_{6,10}]\Delta ^{-1},\nn
P_{1,0}^{((0,0),2,0)}&=&
\chi_{3/80}^{Vir,(4)}\chi_{1/16}^{Vir,(3)}
[\theta_{4,10}^2-\theta_{6,10}^2]
[\theta_{1,10}+\theta_{9,10}]^{-1}\Delta ^{-1},\nn
P_{0,1}^{((1,1),2,0)}&=&
-\chi_{7/16}^{Vir,(4)}[\chi_{\hf }^{Vir,(3)}\theta _{8,10}
                                -\chi_0^{Vir,(3)}\theta_{2,10}]\Delta ^{-1},\nn
P_{0,0}^{((1,1),2,0)}&=&
-\chi_{7/16}^{Vir,(4)}[\chi_0^{Vir,(3)}\theta _{8,10}
                           -\chi_{\hf }^{Vir,(3)}\theta_{2,10}]\Delta ^{-1},\nn
P_{1,0}^{((1,1),2,0)}&=&
-\chi_{7/16}^{Vir,(4)}\chi_{1/16}^{Vir,(3)}
[\theta_{8,10}^2-\theta_{2,10}^2]
[\theta_{3,10}+\theta_{7,10}]^{-1}
\Delta ^{-1},
\eea
where
\ben
\Delta =
\left[(\chi_0^{Vir,(3)})^2-(\chi_{\hf }^{Vir,(3)})^2\right]
[\theta _{2,10}\theta _{4,10} -\theta _{6,10}\theta _{8,10}]\eta ^{-1}.
\een

Two of the six modular forms $F^{\LA }_{2\lp ,n}(q)$ can also be written in
terms of standard modular forms, while the four others are expressed in terms
of the branching functions as announced,
\bea
F^{((0,0),2,0)}_{0,1}(q)
&=&\frac{1}{\chi^{Vir,(3)}_{1/2}(q)}
[P_{0,1}^{((0,0),2,0)}(q)\theta_{4,10}(q)+P_{0,0}^{((0,0),2,0)}(q)
\theta_{6,10}(q)],\nn
F^{((1,1),2,0)}_{0,2}(q)
&=&\frac{1}{\chi^{Vir,(3)}_{1/2}(q)}
[P_{0,1}^{((1,1),2,0)}(q)\theta_{8,10}(q)+P_{0,0}^{((1,1),2,0)}(q)
\theta_{2,10}(q)],\nn
F^{((0,0),2,0)}_{0,2}(q)
&=&\frac{1}{\chi^{Vir,(3)}_{1/2}(q)}
\{ [P_{0,1}^{((0,0),2,0)}(q)\theta_{8,10}(q)+P_{0,0}^{((0,0),2,0)}(q)
\theta_{2,10}(q)]
   -\eta Y^{(4)}_0(q) \},\nn
F^{((0,0),2,0)}_{0,0}(q)
&=&\frac{1}{\chi^{Vir,(3)}_{1/2}(q)}
\{ [P_{0,1}^{((0,0),2,0)}(q)\theta_{0,10}(q)+P_{0,0}^{((0,0),2,0)}(q)
\theta_{10,10}(q)]
   -\eta Y^{(4)}_{3/5}(q) \},\nn
F^{((1,1),2,0)}_{0,1}(q)
&=&\frac{1}{\chi^{Vir,(3)}_{1/2}(q)}
\{ [P_{0,1}^{((1,1),2,0)}(q)\theta_{4,10}(q)+P_{0,0}^{((1,1),2,0)}(q)
\theta_{6,10}(q)]
   +\eta Y^{(4)}_{3/5}(q) \},\nn
F^{((1,1),2,0)}_{0,0}(q)
&=&\frac{1}{\chi^{Vir,(3)}_{1/2}(q)}
\{ [P_{0,1}^{((1,1),2,0)}(q)\theta_{0,10}(q)+P_{0,0}^{((1,1),2,0)}(q)
\theta_{10,10}(q)]
   -\eta Y^{(4)}_0(q) \}.\nn
&&
\eea

\section{A theta function identity.}
\setcounter{equation}{0}
In order to prove the following product identity (cf. eq.(\ref{eq:identity}))
\ben
\theta_{a,1}(q,z_y^2)\theta_{b,3\ktp }(q,z_y^{2/3})=
\sum_{r=0}^{k-1} \theta_{(a+2r)\ktp -b,k\ktp }(q)
\theta_{6r+3a+b,3k}(q,z_y^{2/3}),
\een
where $k=\ktp +3$, simply rewrite the LHS according to the definition
of generalized theta functions, i.e.
\ben
\theta_{a,1}(q,z_y^2)\theta_{b,3\ktp }(q,z_y^{2/3})=
\sum _{n,m}q^{n^2+an+\frac{a^2}{4}+3\ktp m^2 +bm+\frac{b^2}{12 \ktp }}
           z_y^{2n+2\ktp m +a+\frac{b}{3}}.
\een
Rename $2n+2\ktp m=2ks+2r$ with $r=0,...,k-1$, and reorganize the double
sum on $n,m$ by a sum on $s,m$ and $r$ in
such a way that the above expression becomes
\ben
\sum_{r=0}^{k-1}\sum_{s,m}q^{k^2s^2+\ktp (\ktp +3)m^2+sk(2r+a-2\ktp m)
+m(-a\ktp +b-2\ktp r)+r^2+ar+\frac{a^2}{4}+\frac{b^2}{12\ktp }}
z_y^{2ks+2r+a+\frac{b}{3}}.
\een
Finally replace $m \rightarrow -m+s$ in the above sum, and notice that
\ben
r^2+ar+\frac{a^2}{4}+\frac{b^2}{12 \ktp }=
\frac{(a\ktp +2r\ktp -b)^2}{4k\ktp }+\frac{(3a+6r+b)^2}{12k}.
\een

\section{Solution to the master equation }
\setcounter{equation}{0}
In order to determine
the elements $S_{L'}^{~L}$ for $1 \le L \le k-3$ and $0 \le L' \le k-2$,
it is sufficient to consider triality $t=0$
in the master equation  \eqn{master}, and
the representations
$\vec{a}'=(p,p)$, $\vec{a}'=(p,0)$
and $\vec{a}'=(0,p)$
with $p$ integer in the range $0 \le p \le [\frac{\ktp}{3}]$
(in order to have a representation pertaining to the $\epsilon =0$ region).
The entries $S_{L'}^{~0}$ and
$S_{L'}^{~k-2}$ are special but can be obtained along similar lines to those
described below.
As can be seen from the explicit expression \eqn{matrix}, the non-zero
entries of $(M(0)^S)_{\LA}^{~L}$ ($1 \le L \le k-3$) occur for
\bea
&\vec{a}&=(k-2-L,k-2-L-3q)~~~{\rm and}~~~k-1-2L \le 3q \le k-2-L,\\
&\vec{a}&=(k-2-L+3q,k-2-L)~~~{\rm and}~~~L-(k-2) \le 3q \le 2L-(k-1),\\
&\vec{a}&=\left(\hf (L-1+3q),\hf (L-1-3q)\right)~~~{\rm and}~~~
-L+1 \le 3q \le L-1,\\
&\vec{a}&=(L,L-3q)~~~{\rm and}~~~2L-(k-3) \le 3q \le L,\\
&\vec{a}&=(L+3q,L)~~~{\rm and}~~~-L \le 3q \le -2L+(k-3),\\
&\vec{a}&=\left(\hf (k-3-L+3q),\hf (k-3-L-3q)\right)
{}~~~{\rm and}~~~L-(k-3)\le 3q \le
(k-3)-L,\nonumber\\
\eea
and $q$ is further restricted by the condition that $a_1,a_2$ are integers.
According to \eqn{smatrix1},\eqn{smatrix2}, the relevant elements
of the S transform matrix of $\SSS$ characters are given by
\bea
S(t=0)_{\vec{a}}^{(p,p)}
&=&\frac{2}{k\sqrt{3}}\left\{
\sin\left(\frac{2\pi }{k}(p+1)(a_1+1)\right)\right.\nn
&+&\left.\sin\left(\frac{2\pi }{k}(p+1)(a_2+1)\right)
-\sin\left(\frac{2\pi }{k}(p+1)(a_1+a_2+2)\right)\right\},
\eea
\bea
S(t=0)_{\vec{a}}^{(p,0)}&+&S(t=0)_{\vec{a}}^{(0,p)}\nn
=
\frac{2}{k\sqrt{3}}
&\Biggl\{& \sin \left(\frac{2\pi}{k}[(p+1)(a_1+1)-qp]\right)
+\sin\left(\frac{2\pi}{k}[(p+1)(a_2+1)+qp]\right)\nn
&&+ \sin\left(\frac{2\pi}{k}[(a_1+1)+qp]\right)
+\sin \left(\frac{2\pi}{k}[(a_2+1)-qp]\right)\nn
&&-\sin\left(\frac{2\pi}{k}[(a_1+1)+(a_2+1)(p+1)+qp]\right)\nn
&&-\sin\left(\frac{2\pi}{k}[(a_2+1)+(a_1+1)(p+1)-qp]\right)\Biggr \},\nn
\eea
where use has been made of the fact that the relevant representations
$\vec{a}$ should have triality zero, and therefore one can choose
$a_1-a_2=3q$.
When $\vec{a}$ is restricted according to the above, the LHS of the master
equation for the representation $\vec{a}'=(p,p)$ can be written as
\bea
{\rm LHS} &=& -\frac{4\sqrt{3}}{k}\sin\left(\frac{\pi }{k}(p+1)(L+1)\right)
\sum_{6q-3(L+1) \in I}\chi^{3k,S}_{6q-3(L+1)}\nn
&\cdot&\left\{ \cos\left(\frac{\pi }{k}(p+1)(L+1)\right)
-\cos\left(\frac{\pi }{k}(p+1)(6q-3(L+1))\right)
\right\},
\label{eq:lhspp}
\eea
where $I$  symbolizes a sum over the intervals
\bea
I_1&=&I_2=[-2k-L+1,-2k+L-1]\nn
I_3&=&[-L+1,L-1]\nn
I_4&=&I_5=[-2k+L+3,-(L+3)]\nn
I_6&=&[-4k+L+3,-2k-(L+3)].
\eea
The above expression \eqn{lhspp}
is obtained as a sum of six terms corresponding to the
six restrictions on $\vec{a}$ listed previously, where one has relabelled
$3q \equiv 6q'-3(L+1)$ in item 3 in order for $\hf (L-1 \pm 3q)$ to be an
integer, $3q \equiv 6q'+3k-3(L+1)$ in item 6
in order for $\hf (k-3-L+3q)$ to be
an integer, and $q \equiv -q$ in item 5.
The symmetries of the cosine and of the characters is such that one may
replace any interval with minus the same thing
(denoted $\overline{I}_i$) so
that the sum over $I=I_1 \bigcup I_2...\bigcup I_6$
is equivalent to the sum over
\ben
I=I_6\bigcup I_1\bigcup I_4\bigcup I_3\bigcup \overline{I}_5\bigcup
\overline{I}_2,
\een
which in turn is
equivalent to the sum over the interval of length $6k$
\ben
I'=[-4k+(L+3),2k+(L+3)].
\een
Therefore, the LHS \eqn{lhspp} becomes
\bea
{\rm LHS}&=&
-\frac{4\sqrt{3}}{k}\sin\left(\frac{\pi }{k}(p+1)(L+1)\right)
\sum_{q=0}^{k-1}\chi^{3k,S}_{6q-3(L+1)}\nn
&\cdot&\left\{ \cos\left(\frac{\pi }{k}(p+1)(L+1)\right)
-\cos\left(\frac{\pi }{k}(p+1)(6q-3(L+1))\right)
\right\}\nn
&=&-\frac{4\sqrt{3}}{k}\sin\left(\frac{\pi }{k}(p+1)(L+1)\right)
\cdot\frac{1}{\sqrt{6k}}\cdot\sum_{m'\in{\sf Z}_{6k}}\chi^{3k}_{m'}
\sum_{q=0}^{k-1}\cos \left(\frac{\pi}{k}m'(2q-(L+1))\right)\nn
&\cdot&\left\{\cos \frac{\pi}{k}(p+1)(L+1)-
\cos \left(\frac{\pi}{k}(p+1)(6q-3(L+1))\right)\right\}\nn
&=&-\frac{4\sqrt{3}}{k}\sin\left(\frac{\pi }{k}(p+1)(L+1)\right)
\cdot\frac{1}{\sqrt{6k}}\cdot\sum_{m'\in{\sf Z}_{6k}}\nn
&&\left\{\chi^{3k}_{m'}
\cos \left(\frac{\pi}{k}(p+1)(L+1)\right)\right.\nn
&-&\left.\hf (\chi^{3k}_{m'+3(p+1)}+\chi^{3k}_{m'-3(p+1)})
\sum_{q=0}^{k-1}\cos \left(\frac{\pi}{k}m'(2q-(L+1))\right)\right\}\nn
&=&\sqrt{\frac{2}{k}}\sum_{n'=0}^5(-)^{n'L}\nn
&\cdot&\left\{
-\chi_{kn'}^{3k}
\sin \left(\frac{2\pi}{k}(p+1)(L+1)\right)\right.\nn
&+&\left.(\chi^{3k}_{kn'+3(p+1)}+\chi^{3k}_{kn'-3(p+1)})
\sin \left(\frac{\pi}{k}(p+1)(L+1)\right)
\right\},\nn
\label{eq:lhs1}
\eea
where we used the S transform of the ${\cal{A}}_{3k}$ characters in the
following form,
\ben
(\chi^{3k}_m)^S=\frac{1}{\sqrt{6k}}\sum_{m' \in Z_{6k}}\exp\{
\frac{-imm'\pi }{3k}\}
\chi^{3k}_{m'}.
\label{eq:stransf}
\een
Moreover, the RHS of the master equation  when $t=0$ and $\vec{a}'=(p,p)$
is given by
\bea
{\rm RHS} &=&
\sum_{L'}[(M(\epsilon =0))_{(p,p)}^{~L'}
+(M(\epsilon =-1))_{(k-3-2p,p)}^{~L'}
+(M(\epsilon =1))_{(p,k-3-2p)}^{~L'}]S_{L'}^{~L}\nn
&=&\sum_{i=0}^2 \biggl\{\chi^{3k}_{2ik}S_{2p+1}^{~L}
-[\chi^{3k}_{2ik+3(p+1)}+\chi^{3k}_{2ik-3(p+1)}]S_p^{~L}
-\chi^{3k}_{(2i+1)k}S_{k-3-2p}^{~L}\nn
&&~~~~~+[\chi^{3k}_{(2i+1)k+3(p+1)}+\chi^{3k}_{(2i+1)k-3(p+1)}]
S_{k-2-p}^{~L}\biggr\} .
\label{eq:rhs1}
\eea
On the other hand, very similar manipulations for the combined contribution
of \break
$\vec{a}'=(p,0),(0,p)$ lead to the following expression for the LHS
of the master equation ,
\bea
{\rm LHS}&=&
\frac{4\sqrt{3}}{k}\sum_{q=0}^{k-1}\chi_{6q-3(L+1)}^{3k,S}
\left\{
\sin\left(\frac{\pi}{k}(L+1)(p+1)\right)
\cos\left(\frac{\pi}{k}(p+3)(L+1-2q)\right)\right.\nn
&+&
\sin\left(\frac{\pi}{k}(L+1)\right)
\cos\left(\frac{\pi}{k}(2p+3)(L+1-2q)\right)\nn
&-&\left.\sin\left(\frac{\pi}{k}(L+1)(p+2)\right)
\cos\left(\frac{\pi}{k}p(L+1-2q)\right) \right\}\nn
&=&\sqrt{\frac{2}{k}}\sum_{n'=0}^5(-1)^{n'(L+1)}
\left\{
(\chi_{kn'+2p+3}^{3k}+\chi_{kn'-2p-3}^{3k})
\sin\left(\frac{\pi}{k}(L+1)\right)\right.\nn
&+&(\chi^{3k}_{kn'+p+3}+\chi^{3k}_{kn'-p-3})
\sin\left(\frac{\pi}{k}(L+1)(p+1)\right)\nn
&-&\left. (\chi_{kn'+p}^{3k}+\chi_{kn'-p}^{3k})
\sin\left(\frac{\pi}{k}(L+1)(p+2)\right)\right\}.
\label{eq:lhs2}
\eea
A tedious but straightforward analysis of the RHS for $t=0$ and
$\vec{a}'=(0,p),(p,0)$ leads to the following result
\bea
\lefteqn{ {\rm RHS}=
\sum_{i=0}^2\biggl\{ -(\chi_{2ik+(2p+3)}^{3k}+\chi_{2ik-(2p+3)}^{3k})S_0^{~L}+
(\chi_{2ik+p}^{3k}+\chi_{2ik-p}^{3k})S_{p+1}^{~L}}\nn
&+&(\chi_{(2i+1)k+p+3}^{3k}+\chi_{(2i+1)k-p-3}^{3k})S_{k-p-2}^{~L}
-(\chi_{2ik+p+3}^{3k}+\chi_{2ik-p-3}^{3k})S_p^{~L}\nn
&-&(\chi_{(2i+1)k+p}^{3k}+\chi_{(2i+1)k-p}^{3k})S_{k-p-3}^{~L}
+(\chi_{(2i+1)k+2p+3}^{3k}+\chi_{(2i+1)k-2p-3}^{3k})S_{k-2}^{~L}\biggr\}.
\label{eq:rhs2}
\eea
Comparison between \eqn{lhs1} and \eqn{rhs1} on the one hand, and
between \eqn{lhs2} and \eqn{rhs2} on the other
allows us to write
\ben
S_{L'}^{~L}=-\sqrt{\frac{2}{k}}\sin\left(\frac{\pi }{k}(L'+1)(L+1)\right)
\een
for $1 \le L \le k-3$ and $0 \le L' \le k-2$,
which corresponds, up to a sign, to the elements of the S
transform matrix for $\widehat{SU(2)}_{\ktp+1 }$ characters.

\end{document}